\documentclass[journal]{IEEEtran}

\usepackage{ucs}
\usepackage[utf8x]{inputenc}
\usepackage[cmex10]{amsmath}
\usepackage{cite, amsfonts, amssymb, amsthm, bm, bbm, graphicx, relsize, multirow, booktabs, color, siunitx, colortbl}

\usepackage[american]{babel}
\usepackage[T1]{fontenc}
\usepackage{algorithmic, algorithm}

\theoremstyle{remark}
\newtheorem{remark}{Remark}
\newtheorem{example}{Example}

\DeclareMathOperator{\diag}{diag}

\newcommand{\e}{\mathrm{e}}
\renewcommand{\i}{\mathrm{i}}

\newcommand*{\herm}{^{\mathsf{H}}}
\newcommand*{\transp}{^{\mathsf{T}}}

\definecolor{Gray}{gray}{0.9}

\newcommand{\cc}{\cellcolor[gray]{0.9}}

\title{Beampattern Design for Transmit Architectures Based on Reconfigurable Intelligent Surfaces}
\author{Ciro D'Elia, Emanuele~Grossi,~\IEEEmembership{Senior~Member,~IEEE}, Luca~Venturino,~\IEEEmembership{Senior Member,~IEEE} 
\thanks{Copyright \copyright 20xx IEEE. Personal use of this material is permitted. However, permission to use this material for any other purposes must be obtained from the IEEE by sending a request to pubs-permissions@ieee.org.}
\thanks{The authors are with the Department of Electrical and Information Engineering (DIEI), University of Cassino and Southern Lazio, 03043 Cassino, Italy, and with Consorzio Nazionale Interuniversitario per le Telecomunicazioni (CNIT), 43124 Parma, Italy (delia@unicas.it, e.grossi@unicas.it, l.venturino@unicas.it). The work of E. Grossi was supported by the European Union -- Next-GenerationEU -- National Recovery and Resilience Plan (NRRP) -- Mission 4 Component 2, Investment n. 1.3, Call 341 15-03-2022 (Project PE00000001, program ``RESTART,'' CUP n. E63C22002040007). The work of L. Venturino was supported by the European Union -- Next-GenerationEU -- National Recovery and Resilience Plan (NRRP) -- Mission 4 Component 2, Investment n. 1.1, Call PRIN 2022 D.D. 104 02-02-2022 (Project 202238BJ2R CIRCE, CUP n. H53D23000420006).}
}

\begin{document}
\bstctlcite{BSTcontrol}
\maketitle

\begin{abstract}
In this work, we tackle the problem of beampattern design for a transmit system employing a large reconfigurable intelligent surface (RIS) to redirect radio frequency signals emitted by a few active antennas (sources). We begin by establishing a convenient signal model and discussing the impact of signal bandwidth, source-RIS channel, and system geometry on our derivations. Subsequently, we propose a joint optimization of the waveform emitted by each source and the phase shifts introduced by the RIS. The objective is to match a desired space-frequency distribution of the far-field radiation pattern, relevant to both radar and communication applications. We present a sub-optimal solution to this problem, subject to a constraint on the total power radiated by the sources and, optionally, on the constant modulus of the waveforms. The provided example demonstrates the effective beampattern shaping capabilities of this RIS-based transmit architecture. Specifically, for the same array size and the same desired radiation pattern, the resulting approximation error is comparable to that obtained with a fully-digital MIMO array, especially when constant-modulus waveforms are enforced, and significantly smaller than that of a phased array.
\end{abstract}


\section{Introduction}

\IEEEPARstart{T}{he} synthesis of a desired transmit beampattern---a specific spatial and spectral distribution of the amplitude (or power) of the far-field electromagnetic radiation---is a classical problem in digital array processing with broad relevance across numerous wireless applications. For instance, the generation of multiple beampatterns enables crucial functionalities like beam scanning, target tracking, and track-while-scan operations in radar systems~\cite{Blackman_1999}. Similarly, fifth-generation (5G) and beyond (6G) communication systems operating at millimeter-wave (mmWave), sub-terahertz (sub-THz), and terahertz (THz) frequencies leverage directional beams to overcome severe path loss and facilitate multi-user communication through space-division multiple access~\cite{Shafi_2017, Saad_2020, Chaccour_2022}. Furthermore, precisely shaped transmit beampatterns can effectively reduce signal leakage towards potential eavesdroppers and mitigate interference among co-located access points or radar systems~\cite{Jin_2021, Wei_2022, Grossi_2021}. In integrated sensing and communication (ISAC) systems, a dual-functional transceiver can tailor the beampattern to simultaneously perform environmental monitoring and serve communication users~\cite{Hassanien_2016, Johnston_2022, LiuF_2022}, and reconfigurable intelligent surfaces (RISs) offer a promising technology to enhance the flexibility and efficiency of beampattern design in such integrated systems.

Achieving these sophisticated beampatterns often necessitates large (massive) antenna arrays. However, implementing fully digital arrays, where each antenna element is controlled by a dedicated radio frequency (RF) chain, can be prohibitively expensive and energy-intensive~\cite{Gao_2018}. Hybrid analog-digital arrays have been proposed in the past to reduce the number of required RF chains~\cite{Ayach_2014, Molisch_2017}. These architectures connect the outputs of a limited number of RF chains to either all antennas (fully-connected) or a subset thereof (partially-connected) via an analog network. Nevertheless, the scalability of these hybrid approaches remains limited by the power consumption associated with the analog network, particularly as the number of antennas increases~\cite{Yan_2019, Jamali_2021}.

To address these limitations, reconfigurable intelligent surfaces (RISs) have recently emerged as an energy-efficient and cost-effective alternative, finding applications in metasurface-based transmitters~\cite{DiRenzo_2020}, RIS-aided antennas~\cite{Jamali_2021}, and transmitter-type RISs~\cite{Basar_2021}. An RIS is a planar array of tunable elements capable of reflecting incident electromagnetic signals with desired phase and/or amplitude adjustments, without the need for RF chains or significant processing delays. RISs can be passive~\cite{Huang_2007, Hum_2014, Nayeri_2015, He_2019, Tsilipakos_2020} or active~\cite{Larsonn_2021, LiuK_2022, Zhang_2023} and can be controlled and reconfigured almost in real time~\cite{Cui_2014}. They can be either reflecting (where the signal is reflected from the surface) or transmitting (where the signal is transmitted in the forward direction),\footnote{The primary distinction between reflective and transmitting RIS is that the latter uses a pair of patches (one in each half-plane) for receiving and transmitting, whereas the former reuses the same patch for receiving and reflecting~\cite{Jamali_2021, Xie_2023}; nevertheless, both types can accurately be described by the same equivalent lumped circuit model~\cite{Abeywickrama_2020, Dajer_2022}. In general, reflective RISs are less complex (thanks to the easier placement of the control system for the phase shifters on the back side of the surface) and have a larger gain (due to the metal ground plane that reflects the entire incident wave).} but, more recently, simultaneously reflecting and transmitting RISs have also been considered~\cite{Xu_2021}. RISs can be deployed remotely, forming a dynamically controllable smart radio environment~\cite{Wu_2019, DiRenzo_2020, LiuY_2021, Ma_2021, Schroeder_2022, Guo_2023}, or integrated into the transmit architecture, acting as a feed antenna with a fixed and pre-designed source-RIS channel~\cite{Abdelrahman_2017, Pham_2019, Hasani_2019}.

In this work, we focus on the beampattern design for the RIS-based transmit architecture introduced in~\cite{Jamali_2021, Li_2021}. This architecture employs a small number of active antennas, termed sources (potentially a single source, as in~\cite{Li_2021}), each with a dedicated RF chain, to illuminate a passive RIS. The RIS, comprising a large number of low-cost, energy-efficient elements, reflects or retransmits a phase-shifted version of the combined signals from the sources. This approach offers several advantages~\cite{Jamali_2021}. Indeed, the beam-steering capability is directly controlled by the RIS, so the number of active antennas does not have to scale with the number of passive elements in the RIS. Furthermore, unlike hybrid analog-digital arrays, where analog network losses limit scalability for massive MIMO and high-frequency systems~\cite{Abdelrahman_2017}, the wireless source-RIS link, often referred to as a \emph{space feeding mechanism}, provides inherent energy efficiency~\cite{Hum_2014, Nayeri_2015, Abdelrahman_2017}.

The design of the transmit beampattern of an antenna array is a classical problem that has already been studied in the context of MIMO radars, for both narrowband~\cite{Stoica_2007, Fuhrmann_2008, Ahmed_2014, Aubry_2016, Cheng_2017} and broadband~\cite{SanAntonio_2005, He_2011, Aldayel_2017, Alhujaili_2019} array configurations, as well as in biomedical imaging~\cite{Guo_2008}. In these works, the probing waveforms are designed to match a desired beampattern in a least-squares sense. The problem has also been addressed in DFRC systems~\cite{Hassanien_2016, LiuX_2020, Wang_2021, Zhang_2022, Luo_2022}, although typically focusing only on the radar beampattern. To the best of the authors' knowledge, beampattern design for the specific RIS-based transmit architecture considered here has been explored only in~\cite{Rahal_2022, Xiong_2024} for narrowband systems (with a single source in~\cite{Rahal_2022}) and without waveform optimization. With increasing signal bandwidth and/or array size, the need to consider broadband architectures becomes imperative. In such cases, the frequency dependence of the source-RIS channels, RIS element beampattern, and steering vector necessitates the inclusion of the waveforms in the optimization problem, leading to a space-time design. Furthermore, the presence of multiple sources would allow the system to generate complex amplitude distributions across the surface elements through carefully controlled constructive and destructive interference of the emitted signals; this fine-grained control could significantly improve the beam-shaping capability of the system.

In this context, building upon the preliminary findings in~\cite{Grossi_2023}, this work makes the following contributions.\footnote{This work expands upon~\cite{Grossi_2023} by: providing a more detailed system description with a critical discussion of narrowband and broadband operating conditions; considering an additional constant-modulus constraint; presenting analytical derivations of the methods; and offering a more comprehensive numerical analysis.}
\begin{itemize}
 \item We develop a detailed model for the far-field signal of the RIS-based transmit architecture with multiple sources, eliciting the effects of source waveforms, RIS phases, and the source-RIS channels.
 
 \item We discuss the narrowband and broadband operating regimes of this architecture, highlighting the influence of RIS size, source positions, and signal bandwidth.
 
 \item We formulate and present the transmit beampattern design problem for this architecture, jointly optimizing source waveforms and RIS phase shifts under various constraints on the source signals, including the constant-modulus constraint, which is a key requirement for efficiency when saturated power amplifiers are used.
 
 \item We demonstrate that this beampattern design problem, while distinct from those in related works, can be tackled with consolidated methodologies, and we present a sub-optimum solution.
 
 \item Finally, we provide numerical examples showcasing the competitiveness of this architecture compared to fully digital MIMO and hybrid analog-digital MIMO systems in terms of synthesized beampatterns, particularly when the constant-modulus constraint is enforced.
\end{itemize}

Potential applications for this RIS-based transmit architecture include radars (where RIS exploitation has been advocated for performance improvement and cost reduction in~\cite{Buzzi_2021, Aubry_2021, Buzzi_2022, Rihan_2022}), communication and DFRC systems~\cite{Gao_2019, Wang_2019, LiuR_2023} (where selectable beampatterns enable spatial/beam index modulation), and (massive) MIMO communication systems~\cite{Jamali_2021} (for broadcasting control signals). Furthermore, the results presented here are applicable to scenarios where the sources are either in close proximity to the RIS (forming a single physical unit) or located remotely. In the latter case, the sources could even represent distinct cooperative systems (e.g., base stations or radars) dedicating focused beams within specific frequency bands towards the RIS to sense areas or communicate with terminals outside their primary coverage.

The remainder of the paper is organized as follows. Section~\ref{sys_model_sec} describes the considered RIS-based transmit architecture and derives the beampattern. Section~\ref{sys_opt_sec} presents the beampattern matching problem and proposes a suboptimal solution. Section~\ref{perf_an_sec} provides a numerical example to demonstrate the advantages of the proposed approach against standard benchmarks. Finally, Section~\ref{conclusion_sec} concludes the paper and outlines future research directions.

\paragraph*{Notation} In the following, $\mathbb R$ and $\mathbb C$ denote the set of real and complex numbers, respectively, while $\i$ is the imaginary unit; $\Re(a)$ is the real part of $a\in\mathbb C$; $\arg(a)$ is the argument (phase) of $a\in\mathbb C$; vectors and matrices are denoted by lowercase and uppercase boldface letters, respectively; $(\,\cdot\,)^*$, $(\,\cdot\,)\transp$, and $(\,\cdot\,)\herm$ denote conjugate, transpose, and conjugate transpose, respectively; $\bm I_n$ is the $n\times n$ identity matrix; $\bm 1_n$ is the all-one $n$-dimensional column vector; $\diag(\bm a)$ is the diagonal matrix with entries $\bm a$ on the main diagonal; $\odot$ is the Hadamard (element-wise) product; $\Vert \cdot\Vert$ denotes the Euclidean norm; finally, $\mathcal O(\,\cdot\,)$ is the big-O notation, to asymptotically bound a function from above.

\begin{figure*}[t]	
\centering	 \centerline{\includegraphics[width=0.9\textwidth]{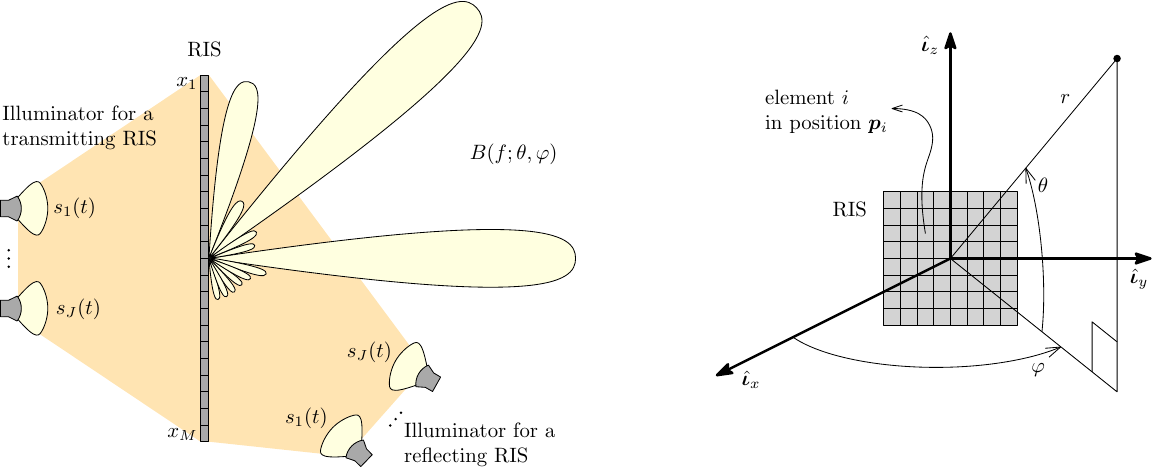}}	
 \caption{Beampattern at frequency $f$ and spatial angle $(\theta,\varphi)$ of a transmit architecture composed of an illuminator with $J$ elements and an RIS with $M$ elements. The illuminator can be located behind the surface (for a transmitting RIS) or in front of it (for a reflecting RIS).} \label{fig_1}
\end{figure*}

\section{System description}\label{sys_model_sec}

In this section, we derive the signal model of the far-field signal of the considered RIS-based transmit architecture, analyze its narrowband and broadband operating regimes (see Sec.~\ref{narrowband_broadband_subsec}), and discuss the case of asynchronous sources (see~Sec.~\ref{asynchronous_subsec}). This detailed signal model is then leveraged by~Sec~\ref{sys_opt_sec} to formulate and (sub-optimally) solve the beampattern design problem.

Consider a transmit architecture composed of an illuminator with $J$ sources and a passive RIS with $M$ elements, as shown in Fig.~\ref{fig_1}. The carrier frequency is $f_c$, and the lowpass signal emitted by the $j$-th source is denoted $s_j(t)$, which is a waveform with support included in the interval $[0,T]$ and Fourier transform approximately equal to zero outside the interval $[-W/2, W/2]$. The $i$-th element of the RIS is located at position $\bm p_i$ and introduces a controllable phase shift. The global reference system is located at the center of gravity of the RIS, whereby $\sum_{i=1}^M \bm p_i = \begin{bmatrix}0 & 0 & 0 \end{bmatrix}\transp$. The frequency response of the channel linking the $j$-th source, the $i$-th element of the RIS, and the point $(r, \theta, \varphi)$ in the far-field region is\footnote{We are assuming here that the possible coupling effects among the element of the RIS and the sources can be neglected.}
\begin{equation}
 H_{ij}(f; r,\theta,\varphi) = \frac{\e^{-2\pi \i f (r/c +\tau_i(\theta,\varphi))}}{\sqrt{4\pi r^2}} \Gamma_i(\theta,\varphi)x_i G_{ij}(f) , \label{channel_expr}
\end{equation}
where $G_{ij}(f)$ is the frequency response of the channel between source $j$ and element $i$ of the RIS, $x_i$ is a unit-modulus scalar modeling the frequency response of the RIS element,\footnote{We are assuming here that $W$ is sufficiently small that the frequency response of the RIS can be considered constant in $[-W/2, W/2]$.} and the remaining terms represent the frequency response of the channel between element $i$ of the RIS and the observation point; in particular, $\Gamma_i(\theta,\varphi)$ is the amplitude beampattern of element $i$ of the RIS in the direction $(\theta,\varphi)$, $\sqrt{4\pi r^2}$ is the term due to the free-space attenuation from the RIS to the observation point, and $r/c +\tau_i(\theta,\varphi)$ is the corresponding propagation delay, with $c$ denoting the speed of light and~\cite[Eq.~(2.16)]{Van_Trees_4}
\begin{equation}
 \tau_i(\theta,\varphi) = -\frac{1}{c} (p_{i,1}\cos\theta \cos\varphi +p_{i,2} \cos \theta \sin \varphi + p_{i,3}\sin \theta) . \label{delays_expr}
\end{equation}

With this notation, the Fourier transform of the complex envelope of the signal observed at $(r,\theta,\varphi)$ is
\begin{align}
 Y(f;r,\theta,\varphi) &= \sum_{i=1}^M \sum_{j=1}^J H_{ij}(f+f_c; r,\theta,\varphi)S_j(f)\notag\\
 &= \frac{\e^{-2\pi \i (f+f_c)r/c}}{\sqrt{4\pi r^2}} \sum_{i=1}^M \e^{-2\pi \i (f+f_c) \tau_i(\theta,\varphi)}  \notag\\
 & \quad \times \Gamma_i(\theta,\varphi) x_i\sum_{j=1}^J G_{ij}(f+f_c) S_j(f) ,\label{Y_expr}
\end{align}
where $S_j(f)$ is the Fourier transform of $s_j(t)$. Letting
\begin{align}
 \bm x & = \begin{bmatrix}x_1 & \cdots & x_M \end{bmatrix}\transp \in \mathbb C^M,\\
 \bm \sigma(f) & = \begin{bmatrix}S_1(f) & \cdots & S_J(f) \end{bmatrix}\transp \in \mathbb C^J,\label{def_sigma}
 \end{align}
and denoting by $\bm \Omega (f; \theta,\varphi)$ the $M\times J$ complex matrix whose entry $(i,j)$ is $\Omega_{ij} (f; \theta,\varphi)= G_{ij}(f+f_c)\Gamma_i(\theta,\varphi)$, we can rewrite~\eqref{Y_expr} in the following compact form
\begin{multline}
 Y(f;r,\theta,\varphi) = \frac{1}{\sqrt{4\pi r^2}}\e^{-2\pi \i (f+f_c)r/c}\\
 \quad \times \bm v\herm(f;\theta,\varphi) \diag (\bm x) \bm \Omega(f; \theta,\varphi) \bm \sigma(f),
\end{multline}
where
\begin{equation}
 \bm v(f;\theta,\varphi) = \begin{bmatrix} \e^{2\pi\i (f+f_c)\tau_1(\theta,\varphi)} & \cdots & \e^{2\pi\i (f+f_c)\tau_M(\theta,\varphi)} \end{bmatrix}\transp \in \mathbb C^M \label{steering_vec}
\end{equation}
is the array manifold vector in the direction $(\theta,\varphi)$ at frequency $f$~\cite[Eq.~(6.669)]{Van_Trees_4}. Therefore, the amplitude beampattern at frequency $f$ and spatial angle $(\theta, \varphi)$ is\footnote{The power spectral density of a (deterministic) signal is obtained by dividing the squared modulus of its Fourier transform (i.e., the energy spectral density) by the signal duration (i.e., $T$). Since we consider the amplitude beampattern, we divide the modulus of the Fourier transform by $\sqrt T$.}
\begin{multline}
 B(f; \theta,\varphi) = \sqrt{\frac{4\pi r^2}{T}}\bigl| Y(f;r,\theta,\varphi)\bigr| \\
 =\frac{1}{\sqrt T} \bigl| \bm v\herm(f;\theta,\varphi) \diag (\bm x) \bm \Omega(f; \theta,\varphi) \bm \sigma(f) \bigr|. \label{ampl_beampattern}
\end{multline}

Since the signals emitted by the sources are bandlimited, i.e., $\bigl|S_j(f) \bigr| \approx 0$ for all $f\notin [-W/2, W/2]$, we have that
\begin{equation}
 S_j(f)=\int_0^T s_j(t) \e^{-2\pi \i ft}dt \approx \frac{1}{W} \sum_{n=1}^N s_j(n/W) \e^{-2\pi \i nf/W}, \label{approx_Sj}
\end{equation}
where $N=\lfloor WT\rfloor$. Therefore, after defining
\begin{align}
 \bm s_n &= \begin{bmatrix}s_1 (n/W) & \cdots & s_J(n/W) \end{bmatrix}\transp \in \mathbb C^J,\\
 \bm s&=\begin{bmatrix}\bm s_1\transp & \cdots & \bm s_N\transp \end{bmatrix}\transp \in \mathbb C^{JN}, \label{s_Def}
\end{align}
and exploiting~\eqref{def_sigma} and~\eqref{approx_Sj}, we can finally approximate~\eqref{ampl_beampattern} as
\begin{align}
& B(f;\theta,\varphi) \approx \frac{1}{W\sqrt T} \biggl| \bm v\herm(f;\theta,\varphi) \diag (\bm x) \bm \Omega(f; \theta,\varphi) \notag\\
 &\quad \times \sum_{n=1}^N \bm s_n \e^{-2\pi \i nf/W} \biggr|
 \notag\\
 &= \frac{1}{W \sqrt T} \Bigl| \bm v\herm(f;\theta,\varphi) \diag (\bm x) \bm \Omega(f; \theta,\varphi) \bigl( \bm e\transp(f) \otimes \bm I_J \bigr) \bm s \Bigr|\notag\\
 &= \bigl| \bm v\herm(f;\theta,\varphi)\diag (\bm x) \bm Q (f;\theta, \varphi) \bm s \bigr|, \label{ampl_beampattern_2}
\end{align}
where
\begin{align}
 \bm e(f)&= \begin{bmatrix}\e^{-2\pi\i f/W} & \cdots & \e^{-2\pi\i N f /W} \end{bmatrix}\transp \in \mathbb C^{N},\\
 \bm Q (f; \theta, \varphi)& = \frac{1}{W \sqrt T} \bm \Omega(f; \theta,\varphi) ( \bm e\transp(f) \otimes \bm I_J ) \in \mathbb C^{M\times JN} . 
\end{align}
Observe that~\eqref{ampl_beampattern} and~\eqref{ampl_beampattern_2} represent an extension to the considered RIS-based architecture of the amplitude beampattern for MIMO systems analyzed in~\cite{He_2011}.

\subsection{Narrowband/broadband architectures} \label{narrowband_broadband_subsec}

In general, the channel between source $j$ and element $i$ of the RIS can be modeled as a tap-delay line, i.e.,
\begin{equation}
 G_{ij}(f)\approx \sum_{\ell=1}^{L_{ij}} g_{ij\ell} \e^{-2\pi \i f t_{ij\ell}},
\end{equation}
where $L_{ij}$ is the number of taps, $g_{ij\ell}$ and $t_{ij\ell}$ are the attenuation and delay of the $\ell$-th tap, respectively, and $\max_\ell t_{ij\ell} - \min_\ell t_{ij\ell}$ is the delay spread. Upon defining
\begin{subequations}
 \begin{align}
 \delta & =\frac12 \left( \max_{i,j,\ell} t_{ij\ell} + \min_{i,j,\ell} t_{ij\ell}\right),\\
 \delta_{ij\ell}& = t_{ij\ell}- \delta, \quad \forall i,j,\ell,
 \end{align}%
\end{subequations} 
the complex envelope in~\eqref{Y_expr} can be rewritten as
\begin{align}
 Y(f;r,\theta,\varphi) & \approx 
 \frac{\e^{-2\pi \i (f+f_c)(r/c+\delta)}}{\sqrt{4\pi r^2}} \sum_{i=1}^M \e^{-2\pi \i f_c \tau_i(\theta,\varphi)} \notag \\
 & \quad \times \Gamma_i(\theta,\varphi) x_i \sum_{j=1}^J \sum_{\ell=1}^{L_{ij}} g_{ij\ell}\e^{-2\pi \i f_c\delta_{ij\ell}}\notag\\
 & \quad \times S_j(f) \e^{-2\pi \i f (\tau_i(\theta,\varphi)+\delta_{ij\ell})}.\label{Y_expr_multipath}
\end{align}
If the delay offset $|\tau_i (\theta,\varphi) + \delta_{ij\ell}|$ is sufficiently small to have \begin{equation}\label{eq_approx_narrowband}
\e^{-2\pi \i f (\tau_i(\theta,\varphi)+\delta_{ij\ell})}\approx 1,\quad \forall f\in[f_c-W/2, f_c+W/2],
\end{equation}
then~\eqref{Y_expr_multipath} can further be approximated as
\begin{multline}
 Y(f;r,\theta,\varphi) \approx \frac{\e^{-2\pi \i (f+f_c)(r/c+\delta)}}{\sqrt{4\pi r^2}} \sum_{i=1}^M \e^{-2\pi \i f_c \tau_i(\theta,\varphi)} \\
 \times\Gamma_i(\theta,\varphi) x_i \sum_{j=1}^J \left(\sum_{\ell=1}^{L_{ij}} g_{ij\ell}\e^{-2\pi \i f_c\delta_{ij\ell}} \right) S_j(f). \label{Y_narrowband}
\end{multline}
Notice that, in the time-domain,~\eqref{Y_narrowband} becomes
\begin{multline}
 y(t;r,\theta,\varphi) \approx \frac{\e^{-2\pi \i f_c(r/c+\delta)}}{\sqrt{4\pi r^2}}   \sum_{i=1}^M \e^{-2\pi \i f_c \tau_i(\theta,\varphi)} \Gamma_i(\theta,\varphi) x_i\\
 \times\sum_{j=1}^J \left(\sum_{\ell=1}^{L_{ij}} g_{ij\ell} \e^{-2\pi \i f_c\delta_{ij\ell}}\right) s_j\bigl(t- r/c -\delta\bigr), \label{y_complex_envelop_narrowband}
\end{multline}
which shows that the signals $\{s_j(t)\}_{j=1}^J$ are now all delayed by the same quantity $r/c +\delta$ at the observation point and that the delay differences $\tau_i(\theta,\varphi)+\delta_{ij\ell}$ simply reduce to phase shifts, for all $\theta,\varphi,i,j,\ell$.

The approximations in~\eqref{eq_approx_narrowband}, \eqref{Y_narrowband}, and~\eqref{y_complex_envelop_narrowband} are valid if\begin{multline}
 W \max_{\theta, \varphi} \left\{\max_i \left(\tau_i(\theta,\varphi) + \max_{j,\ell} \delta_{ij\ell} \right)\right.\\
 \left.- \min_i \left( \tau_i(\theta,\varphi) + \min_{j,\ell} \delta_{ij\ell}\right) \right\} \ll 1.\label{narrowband_condition}
\end{multline}
If~\eqref{narrowband_condition} is satisfied, we say that the RIS-based architecture is \emph{narrowband}; otherwise, we say that it is \emph{broadband}. Notice that~\eqref{narrowband_condition} is a generalization of the well-known narrowband array condition~\cite[Eq.~(2.47)]{Van_Trees_4} to the considered architecture, as it also accounts for the propagation delay offsets $\{\delta_{ij\ell}\}_{ij\ell}$ along the source-RIS channels.

\begin{example} Consider a square planar RIS, with element field-of-view of $\SI{120}{\degree}$, illuminated by four point-like sources emitting signals with bandwidth $W=\SI{75}{\MHz}$; the sources are located at the same distance from the RIS in correspondence of its four corners, and $G_{ij}(f)$ contains only the line-of-sight path, for all $i,j$. Then, the left-hand side of~\eqref{narrowband_condition} is equal to $0.49$, if the sources-RIS distances and the side length of the RIS are \SI{1}{\m}, while it is equal to 0.098, if they are reduced to \SI{20}{\cm}: in the former case, the system is broadband, while in the latter it is narrowband.
\end{example}

\begin{remark}
When the sources are closely spaced, a good rule of thumb for the narrowband condition can be found by upper bounding the left-hand side of~\eqref{narrowband_condition}. To this end, let $\Delta_\text{RIS}$ and $\Delta_\text{sources}$ be the diameters of the smallest balls circumscribing the RIS and the sources, respectively. Then
\begin{align}
 \max_{\theta, \varphi} & \left\{\max_i \left(\tau_i(\theta,\varphi) + \max_{j,\ell} \delta_{ij\ell}\right) \right.\notag\\
 & \quad \left.- \min_i\left( \tau_i(\theta,\varphi) + \min_{j,\ell}\delta_{ij\ell}\right) \right\} \notag\\
 & \leq \max_{\theta, \varphi} \left(\max_i \tau_i(\theta,\varphi) - \min_i \tau_i(\theta,\varphi)\right) \notag\\
 & \quad + \max_{i,j,\ell}\delta_{ij\ell} - \min_{i,j,\ell} \delta_{ij\ell}\notag\\
 & \leq \frac{\Delta_\text{RIS}}{c} + \frac{\Delta_\text{RIS}+ \Delta_\text{sources}}{c}, \label{upper_bound_delays}
\end{align}
and\footnote{Observe that the first inequality in~\eqref{upper_bound_delays} comes from the fact that the maximum (minimum) of a sum is always smaller (greater) than the sum of the maxima (minima), while the second inequality is an upper bound obtained when the wave is parallel to the RIS, and sources and RIS lie in the same plane, whereby $\max_{\theta,\varphi} \bigl( \max_i \tau_i(\theta,\varphi) - \min_i \tau_i(\theta,\varphi)\bigr) =\Delta_\text{RIS}/c$ and $\max_{i,j}\delta_{ijL_{ij}} - \min_{i,j} \delta_{ij1} =(\Delta_\text{RIS}+\Delta_\text{sources})/c$.} a sufficient condition for the narrowband condition in~\eqref{narrowband_condition} to hold is\footnote{Notice that $c/W$ is the approximate range resolution of a waveform with bandwidth $W$.}
\begin{equation}
 2\Delta_\text{RIS} + \Delta_\text{sources} \ll c/W . \label{narrowband_cond_2}
\end{equation}
This can best be appreciated if compared with the equivalent sufficient condition for MIMO systems, namely, $\Delta_\text{MIMO}\ll c/W$, where $\Delta_\text{MIMO}$ is the diameter of the smallest ball circumscribing the fully digital array; indeed,~\eqref{narrowband_cond_2} accounts now for the presence of two arrays (the RIS and the sources) and for the fact that the wave travels first from the source to the RIS and then from the RIS to the observation point.
\end{remark}

By comparing~\eqref{Y_narrowband} with~\eqref{Y_expr}, we see that, in the narrowband case, the amplitude beampattern retains the same expression as in~\eqref{ampl_beampattern} and~\eqref{ampl_beampattern_2}, but the array manifold vector and the source-RIS channels are now frequency independent; namely, $\bm v(f;\theta,\varphi)$ becomes
\begin{equation}
 \bm v(\theta,\varphi) = \begin{bmatrix} \e^{2\pi\i f_c\tau_1(\theta,\varphi)} & \cdots & \e^{2\pi\i f_c\tau_M(\theta,\varphi)} \end{bmatrix}\transp,
\end{equation}
and $\bm \Omega (f;\theta,\varphi)$ becomes $\bm \Omega (\theta,\varphi)$, with entry $(i,j)$ given by
\begin{equation}
 \Omega_{ij}(\theta, \varphi) =\Gamma_i(\theta,\varphi) \sum_{\ell=1}^{L_{ij}} g_{ij\ell} \e^{-2\pi \i f_c\delta_{ij\ell}}.
\end{equation}
In this case, the dependency of the power distribution from the frequency is determined solely by the source signals and is often removed by integration, i.e.,
\begin{align}
 & \bar B^2 (\theta,\varphi) = \int_{-W/2}^{W/2} B^2(f;\theta,\varphi)df \notag\\
 & \quad = \bm v\herm (\theta,\varphi) \diag(\bm x) \bm \Omega(\theta, \varphi) \left(\frac{1}{T} \int_{-W/2}^{W/2} \bm \sigma(f) \bm \sigma\herm(f) df\right) \notag \\
 & \qquad \times \bm \Omega\herm(\theta, \varphi)\diag(\bm x^*) \bm v (\theta,\varphi). 
\end{align}
Since
\begin{align}
 & \left(\frac{1}{T} \int_{-W/2}^{W/2} \bm \sigma(f) \bm \sigma\herm(f) df\right)_{hk} = \frac 1T \int_{-W/2}^{W/2} S_h(f) S_k^*(f) df \notag\\
 &\quad = \frac 1T \int_0^T s_h(t) s_k^*(t) dt\notag \\
 & \quad \approx \frac{1}{WT} \sum_{n=1}^N s_h\left(\frac nW\right) s_k^*\left( \frac nW\right) \notag\\
 & \quad \approx \left(\frac{1}{N} \sum_{n=1}^N \bm s_n \bm s_n^H\right)_{hk}, \quad \forall h,k,
\end{align}
where the last approximation follows from the fact that $N=\lfloor WT\rfloor$, we have that
\begin{align}
 \bar B^2(\theta,\varphi) &= \bm v\herm (\theta,\varphi) \diag(\bm x) \bm \Omega(\theta, \varphi) \left(\frac 1N \sum_{n=1}^N \bm s_n \bm s_n\herm \right) \notag \\
 &\quad \times \bm \Omega\herm(\theta, \varphi)\diag(\bm x^*) \bm v (\theta,\varphi). \label{pwoer_beampattern_angle}
\end{align}
Observe that~\eqref{pwoer_beampattern_angle} represents an extension to the considered RIS-based architecture of the power beampattern over the space domain for MIMO systems analyzed in~\cite{Stoica_2007}.

\subsection{Asynchronous sources}\label{asynchronous_subsec}

When source signal $j$ has support in $[\Delta_j, \Delta_j+T]$, where $\Delta_j\geq0$ accounts for some initial transmission offset, we have 
\begin{align}
 S_j(f)&=\int_{\Delta_j}^{\Delta_j+T} s_j(t) \e^{-2\pi \i ft}dt \notag\\
&= \e^{-2\pi \i f\Delta_j} \int_{0}^{T} s_j(\Delta_j+t) \e^{-2\pi \i ft}dt
\notag\\
&\approx \frac{ \e^{-2\pi \i f\Delta_j}}{W} \sum_{n=1}^N s_j(\Delta_j+n/W) \e^{-2\pi nf/W},
\end{align}
for $j=1,\ldots,J$. In this case, our model still remains valid upon absorbing the phase term $\e^{-2\pi \i f\Delta_j}$ into the channel $G_{ij}(f)$ and defining $\bm s_n$ in~\eqref{s_Def} as 
\begin{equation}
 \bm s_n= \begin{bmatrix}s_1 (\Delta_1+n/W) & \cdots & s_J(\Delta_J+n/W) \end{bmatrix}\transp \in \mathbb C^J,
\end{equation}
for $n=1,\ldots,N$. For example, introducing some delay offsets may be desirable when the sources are located at different distances from the center of gravity of the RIS; in particular, closer sources may be delayed to ensure that all signals simultaneously reach the center of gravity of the RIS.

\section{System Optimization} \label{sys_opt_sec}

The objective here is to design the signal emitted by the illuminator $\bm s$ and the RIS phases $\bm x$ in such a way that the amplitude beampattern $B(f;\theta, \varphi)$ matches in a least squares (LS) sense the desired amplitude beampattern, say $D(f; \theta, \varphi)$. The LS amplitude beampattern matching is a well-accepted design criterion that has been extensively used in recent years (see, e.g.,~\cite{He_2011}). All following derivations rely on the general expression of the amplitude beampattern given in~\eqref{ampl_beampattern_2}, whereby they apply to both broadband and narrowband RIS-based architectures. 

To proceed, define the relative square error (RSE) between the desired beam pattern and the actual beam pattern over the angular and frequency space as
\begin{align}
 \text{RSE}(\bm s, \bm x) &= \frac{1}{4\pi} \int_{-W/2}^{W/2} \int_{-\pi/2}^{\pi/2}\int_{-\pi/2}^{\pi/2} \tilde w(f;\theta,\varphi) \notag \\
 &\quad \times \bigl( D(f;\theta, \varphi) - B(f;\theta, \varphi)\bigr)^2 \cos \theta d\theta d\varphi df \notag\\
 &\quad \times \Bigg(\frac{1}{4\pi} \int_{-W/2}^{W/2} \int_{-\pi/2}^{\pi/2}\int_{-\pi/2}^{\pi/2} \tilde w(f;\theta,\varphi) \notag \\
 &\quad \times D^2(f;\theta, \varphi) \cos \theta d\theta d\varphi df\Bigg)^{-1},
\end{align}
where $\tilde w(f;\theta,\varphi)$ is a weighting function that allows for the emphasis of different regions; observe that, if $\tilde w(f;\theta,\varphi)=1$, the denominator is simply the power radiated with the desired beampattern. If we discretize the angular region $[-\pi/2, \pi/2]^2$ using a rectangular grid with $L$ points uniformly-spaced along each dimension, namely $\{ (\theta_\ell, \varphi_\ell)\}_{\ell=1}^L$, and the frequency region $[-W/2, W/2]$ using $K$ uniformly-spaced points, namely $\{f_k\}_{k=1}^K$, the RSE can be approximated as
\begin{multline}
 \text{RSE}(\bm s, \bm x) \approx \sum_{k=1}^K \sum_{\ell=1}^L w (f_k;\theta_\ell, \varphi_\ell) \\
\quad \times \bigl( D(f_k;\theta_\ell, \varphi_\ell) - B(f_k;\theta_\ell, \varphi_\ell)\bigr)^2,
\end{multline}
where
\begin{align}
 w(f;\theta,\varphi) &= \frac{W \pi^2}{K L}\tilde w(f;\theta,\varphi) \cos \theta\notag\\
 &\quad \times \Bigg( \int_{-W/2}^{W/2} \int_{-\pi/2}^{\pi/2}\int_{-\pi/2}^{\pi/2} \tilde w(f';\theta',\varphi') \notag \\
 &\quad \times d^2(f';\theta', \varphi') \cos \theta d\theta' d\varphi' df'\Bigg)^{-1}. \label{weights_expr}
\end{align}
Hence, the beampattern matching problem tackled here is
\begin{equation}
 \begin{aligned}
 \min_{\bm s \in \mathbb C^{JN}, \bm x \in \mathbb C^M } & \; \text{RSE}(\bm s, \bm x),\\
 \text{s.t.} & \; \frac{1}{N} \Vert \bm s \Vert^2 \leq P, \\
 & \; |x_i|=1, \quad i=1,\ldots,M,
 \end{aligned} \label{problem_1}
\end{equation}
where $P$ is the total power available at the illuminator.\footnote{We have chosen here a \emph{global} power constraint, but individual power constraints (more suited when the sources are widely spaced from each other and from the RIS) can similarly be handled. In this case, $N^{-1}\Vert \bm s \Vert^2\leq P$ would be replaced by $N^{-1} \sum_{n=1}^N |s_{j+(n-1)J}|^2\leq P_j$, $j=1,\ldots,J$, where $P_j$ is the available power at source $j$.} This problem differs from that usually considered in MIMO radars since the beampattern is not only a function of the steering vector (from the array to the observation point) but also of the source-RIS channels, and the degrees of freedom are not only the waveforms radiated by the active sources but also include the RIS phases. For notational convenience, we define
\begin{subequations}
\begin{align}
 w_{k\ell}&= w(f_k;\theta_\ell,\varphi_\ell), & D_{k\ell}&= D(f_k;\theta_\ell,\varphi_\ell),\\
 \bm v_{k\ell}&= \bm v(f_k;\theta_\ell,\varphi_\ell), & \bm Q_{k\ell} &= \bm Q(f_k;\theta_\ell,\varphi_\ell),
\end{align}%
\end{subequations}
for $k=1,\ldots,K$ and $\ell=1,\ldots,L$, so that Problem~\eqref{problem_1} becomes
\begin{equation}
 \begin{aligned}
 \min_{\bm s \in \mathbb C^{JN}, \bm x \in \mathbb C^M} & \; \sum_{k=1}^K \sum_{\ell=1}^L w_{k\ell} \bigl( D_{k\ell} - | \bm v_{k\ell}\herm \diag (\bm x) \bm Q_{k\ell} \bm s | \bigr)^2,\\
 \text{s.t.} & \; \Vert \bm s \Vert^2 \leq N P, \\
 & \; |x_i|=1, \quad i=1,\ldots,M.
 \end{aligned}\label{problem_2}
\end{equation}

In the following, we describe a possible (sub-optimal) procedure to solve this problem. The case where a constant-modulus constraint is imposed on the illuminator signals is instead tackled in Sec.~\ref{constr_s_sec}.

\subsection{Sub-optimum solution to Problem~\eqref{problem_2}}\label{min_prob_general}

Problem~\eqref{problem_2} is intractable due to the non-differentiability of the objective function (caused by the absolute value), and a common approach~\cite{He_2011} is to exploit the fact that, for any $a>0$ and $b\in\mathbb C$,
\begin{equation}
 \min_{\psi \in \mathbb R} | a \e^{\i\psi} - b |^2 = \bigl( a - | b| \bigr)^2, \text{ for } \psi=\arg(b). \label{trick}
\end{equation}
As a consequence, upon introducing the auxiliary variables $\{\psi_{k\ell}\}$, Problem~\eqref{problem_2} can be equivalently rewritten as
\begin{equation}
 \begin{aligned}
 \min_{\substack{\bm s \in \mathbb C^{JN}, \bm x \in \mathbb C^M,\\ \{\psi_{k\ell}\}\in \mathbb R^{KL}}} & \; \sum_{k=1}^K \sum_{\ell=1}^L w_{k\ell} \left| D_{k\ell}\e^{\i\psi_{k\ell}} - \bm v_{k\ell}\herm \diag (\bm x) \bm Q_{k\ell} \bm s\right|^2,\\
 \text{s.t.} & \; \Vert \bm s \Vert^2 \leq N P, \\
 & \; |x_i|=1, \quad i=1,\ldots,M.
 \end{aligned} \label{problem_3}
\end{equation}

This problem is still very complex due to the non-convexity of both the objective function and the constraint set, whereby we resort to sub-optimal procedures to solve it. In particular, we tackle it by resorting to the block-coordinate descent method~\cite{Bertsekas_1999}, also known as non-linear Gauss-Seidel method or as alternating minimization: starting from a feasible point, the objective function is minimized with respect to each of the ``block coordinate'' variables, taken in cyclic order while keeping the other ones fixed at their previous values.

The block-coordinate variables that we consider here are the auxiliary variables $\{\psi_{k\ell}\}_{k,\ell}$, the illuminator signals $\bm s$, and the RIS phases $x_1,\ldots,x_M$. The auxiliary variables can be disjointly minimized, and, from~\eqref{trick}, the solution is
\begin{equation}
\psi_{k\ell}=\arg \bigl(\bm v_{k\ell}\herm \diag (\bm x) \bm Q_{k\ell} \bm s \bigr), \label{sol_sub_prob_psi}
\end{equation}
for $k=1,\ldots,K$ and $\ell=1,\ldots,L$.

As to the illuminator signals, upon defining
\begin{subequations} \label{def_A_b}
\begin{align}
 \bm A & = \sum_{k=1}^K \sum_{\ell=1}^L w_{k\ell} \bm Q_{k\ell}\herm \diag(\bm x^*)\bm v_{k\ell} \bm v_{k\ell}\herm \diag (\bm x) \bm Q_{k\ell} \label{mat_A_def} ,\\
 \bm b &= \sum_{k=1}^K \sum_{\ell=1}^L w_{k\ell} D_{k\ell} \e^{\i \psi_{k\ell}} \bm Q_{k\ell}\herm \diag(\bm x^*) \bm v_{k\ell} ,
\end{align}%
\end{subequations}
the sub-problem to be solved can be equivalently written as
\begin{equation}
 \begin{aligned}
 \min_{\bm s \in \mathbb C^{JN}} & \; \bigl\{ \bm s\herm \bm A \bm s - 2 \Re (\bm s\herm \bm b) \bigr\} ,\\
 \text{s.t.} & \; \Vert \bm s \Vert^2 \leq N P .
 \end{aligned} \label{sub_prob_s_2}
\end{equation}
This is a standard problem, and a solution can be found in terms of a diagonalization of $\bm A$. Specifically, let $\bm U \bm \Sigma \bm U\herm$ be the eigenvalue decomposition of $\bm A$, where $\bm U\in \mathbb C^{JN\times JN}$ is unitary, and $\bm \Sigma= \diag (\sigma_1 \cdots \sigma_{JN})$, with $\sigma_i\geq0$, for all $i$; also, let $\tilde{\bm s}= \bm U\herm \bm s$ and $\tilde{\bm b}= \bm U\herm \bm b$. Then,
\begin{equation}
 \bm s = \bm U \tilde{\bm s}\label{s_update}
\end{equation}
is a solution to Problem~\eqref{sub_prob_s_2} if $\tilde{\bm s}$ is chosen as follows.
\begin{itemize}
 \item If $\sum_{i=1, \sigma_i\neq 0}^{JN} |\tilde b_i|^2/\sigma_i^2 > NP$ or $\tilde b_i\neq 0$ for some $i$ such that $\sigma_i=0$, then $\tilde s_i= \tilde b_i/(\sigma_i+\lambda)$, for all $i$, where $\lambda >0$ is the unique solution of\footnote{Notice that upper and lower bound to $\lambda$ are available (see the Appendix), and the solution can easily be found with the bisection algorithm since the left-hand side of~\eqref{impicit_eq} is continuous, strictly decreasing, and strictly convex for $\lambda>0$.}
 \begin{equation}
 \sum_{i=1}^{JN} \frac{|\tilde b_i|^2}{(\sigma_i+\lambda)^2} = NP. \label{impicit_eq}
 \end{equation}
 \item Otherwise, $\tilde s_i= \tilde b_i/\sigma_i$ for all $i$ such that $\sigma_i\neq 0$, and $\tilde s_i= 0$, for all $i$ such that $\sigma_i= 0$. 
\end{itemize}
The proof is very simple and is reported in the Appendix. 

Let us finally tackle the minimization over the RIS phases. Since $\diag (\bm x) \bm Q_{k\ell} \bm s = \diag (\bm Q_{k\ell} \bm s) \bm x$, upon defining
\begin{subequations} \label{def_B_c}
\begin{align}
 \bm B & = \sum_{k=1}^K \sum_{\ell=1}^L w_{k\ell} \diag(\bm Q_{k\ell} \bm s)^*\bm v_{k\ell} \bm v_{k\ell}\herm \diag (\bm Q_{k\ell} \bm s) , \label{matrix_B_def}\\
 \bm c &= \sum_{k=1}^K \sum_{\ell=1}^L w_{k\ell} D_{k\ell} \e^{\i \psi_{k\ell}} \diag(\bm Q_{k\ell} \bm s)^* \bm v_{k\ell} ,
\end{align}%
\end{subequations}
the sub-problem to be solved can be equivalently written as
\begin{equation}
 \min_{x_i \in \mathbb C: |x_i|=1} \bigl\{ \bm x\herm \bm B \bm x - 2 \Re (\bm x\herm \bm c) \bigr\}, \label{sub_prob_x}
\end{equation}
and its solution is
\begin{equation}
 x_i= \frac{c_i-\sum_{j=1, j\neq i}^M B_{ij} x_j}{\bigl| c_i-\sum_{j=1, j\neq i}^M B_{ij} x_j\bigr|}. \label{xi_update}
\end{equation}

\begin{algorithm}[t] 
\caption{Sub-optimal solution to Problem~\eqref{problem_3}\label{block-coordinate_descent_alg}}
 \begin{algorithmic}[1]
 \renewcommand{\algorithmicrequire}{\textbf{Input:}}
 \renewcommand{\algorithmicensure}{\textbf{Output:}}
 \REQUIRE $\{w_{k\ell}, D_{k\ell}, \bm v_{k\ell}, \bm Q_{k\ell}\}_{k,\ell}$
 \STATE choose $\bm s \in \mathbb C^{JN}: \Vert \bm s \Vert^2\leq NP$
 \STATE choose $\bm x \in \mathbb C^M: |x_i|=1$ $\forall i$
 \STATE update $\{\{\psi_{k\ell}\}_{k=1}^K\}_{\ell=1}^L$ with~\eqref{sol_sub_prob_psi}
 \REPEAT
 \STATE compute $\bm A$ and $\bm b$ in~\eqref{def_A_b}
 \STATE compute the eigenvalue decomposition of $\bm A$
 \STATE update $\bm s$ with~\eqref{s_update}
 \STATE compute $\bm B$ and $\bm c$ in~\eqref{def_B_c}
 \FOR{$i=1,\ldots,M$}
  \STATE update $x_i$ with~\eqref{xi_update}
 \ENDFOR
 \STATE update $\{\{\psi_{k\ell}\}_{k=1}^K\}_{\ell=1}^L$ with~\eqref{sol_sub_prob_psi}
 \UNTIL convergence
 \ENSURE $\bm s$ and $\bm x$
 \end{algorithmic}
\end{algorithm}

The complete routine is reported in Alg.~\ref{block-coordinate_descent_alg}. Notice that each sub-problem in the block-coordinate algorithm is optimally solved, whereby the objective function in different iterations is non-increasing; since the objective function is also non-negative (it is a squared error), Alg.~\ref{block-coordinate_descent_alg} converges. However, since Problem~\eqref{problem_3} is not convex, and the feasible set cannot be expressed as the Cartesian product of closed convex sets, there is no guarantee that a global minimum is reached.

\subsection{Constant-modulus illuminator signals}\label{constr_s_sec}

In Problem~\eqref{problem_2}, there is an overall power constraint on the illuminators. In many cases, however, the source signals are required to be constant-modulus, and only the power allocation among the illuminators can be optimized. In this situation, 
\begin{align}
 \bm s & = \diag (\bm \omega) (\bm 1_N \otimes \bm I_J) \bm u = \bigl(\bm I_N \otimes \diag (\bm u)\bigr) \bm \omega\notag\\
 & = (\bm 1_N \otimes \bm u) \odot \bm \omega,
\end{align}
where $\bm u\in\mathbb R^J$ is the amplitude vector of the illuminator signals, and $\bm \omega \in \mathbb C^{JN}$ is the vector containing the unit-modulus complex scalars accounting for the phase terms, and Problem~\eqref{problem_2} is recast as
\begin{equation}
 \begin{aligned}
  \min_{\substack{\bm u\in\mathbb R^J, \bm \omega \in \mathbb C^{JN}, \\\bm s \in \mathbb C^{JN}, \bm x \in \mathbb C^M}} & \; \sum_{k=1}^K \sum_{\ell=1}^L w_{k\ell} \bigl( D_{k\ell} - | \bm v_{k\ell}\herm \diag (\bm x) \bm Q_{k\ell} \bm s | \bigr)^2,\\
 \text{s.t.} & \; \bm s = (\bm 1_N \otimes \bm u) \odot \bm \omega,\\
  & \; \Vert \bm u \Vert^2 \leq P,\\
  & \; | \omega_i | = 1, \quad i=1,\ldots,JN,\\
  & \; |x_i|=1, \quad i=1,\ldots,M.
 \end{aligned}\label{problem_constant_mod}
\end{equation}

This problem can (sub-optimally) be solved by introducing the auxiliary variables $\{\psi_{k\ell}\}_{k,\ell}$ and then adopting the block-coordinate descent algorithm in Sec.~\ref{min_prob_general}, where the block-coordinate variables are $\{\psi_{k\ell}\}_{k,\ell}$, $\bm u$, $\omega_1,\ldots, \omega_{JN},x_1,\ldots,x_M$.

\section{Performance analysis} \label{perf_an_sec}

In this section, we present a numerical example to demonstrate the advantages of the RIS-based transmit architecture shown in Fig.~\ref{fig_1} compared to potential competitors. The simulation setup is described in the next section. In Sec.~\ref{sec_competitive_systems}, we introduce the benchmark systems and their corresponding beampattern optimization problems. Sec.~\ref{complexity_Sec} analyzes the computational complexity required for solving the beampattern optimization problem for both the RIS-based architecture and the benchmark systems. Finally, Sec.~\ref{num_res_sec} presents and discusses the simulation results.

\subsection{System settings}

We consider a $10 \times 10$ transmitting RIS, so that $M=100$, illuminated by $J=4$ sources. The signals have duration $T=\SI{0.64}{\us}$ and bandwidth $W=\SI{100}{\MHz}$, so that $N=\lfloor WT\rfloor=64$. 
The carrier frequency is $f_c=\SI{3}{\GHz}$, and the inter-element spacing of the RIS is half-wavelength of the highest frequency, i.e., $c/\bigl(2(f_c+W/2)\bigr)\approx\SI{4.915}{\cm}$. The 4 sources are placed at a distance of \SI{60}{\cm} from the RIS, each in correspondence with one of the four quadrants of the RIS. All the elements of the RIS and of the illuminator have a cosine power beampattern, and

\begin{align}
 G_{ij}(f) & =  \underbrace{\frac{\e^{-2\pi \i fd_{ij}/c}}{\sqrt{4\pi d_{ij}^2}}}_\text{(a)} \biggl(\underbrace{2(\beta+1) \cos^\beta \theta_{j\rightarrow i} \cos^\beta \varphi_{j\rightarrow i}}_\text{(b)} \notag\\
  & \quad \times \underbrace{\frac{(c/f)^2}{4\pi} 4 \cos \theta_{i\rightarrow j} \cos \varphi_{i\rightarrow j}}_\text{(c)} \; \biggr)^{1/2}, \label{channel_freq_resp}\\
  \Gamma_i(\theta,\varphi) & =(4\cos \theta \cos \varphi)^{1/2},
\end{align}
where: (a) accounts for the free-space propagation, with $d_{ij}$ the distance between the $i$-th element of the RIS and the $j$-th source; (b) is the gain of the $j$-th source towards the $i$-th element of the RIS, with $\beta=13$, and $\theta_{j\rightarrow i}$ and $ \varphi_{j\rightarrow i}$ the elevation and azimuth angles of element $i$ of the RIS as seen from element $j$ of the illuminator, respectively; and (c) is the effective aperture of the $i$-th element of the RIS from the direction of the $j$-th source, with $\theta_{i\rightarrow j}$ and $ \varphi_{i\rightarrow j}$ the elevation and azimuth angles of element $j$ of the illuminator as seen from element $i$ of the RIS, respectively. Note that, while individual elements of the illuminator and the RIS are within each other's far-field, the arrays as a whole are not, whereby the channel $G_{ij}(f)$ describes a spherical-wave propagation.

The desired beampattern is composed of a broad beam for the negative frequencies and a narrow beam for all frequencies, i.e.,
\begin{equation}
 D(f;\theta, \varphi) =\begin{cases}
 \frac{16}{\sqrt{W \left(\sqrt 2 - \sin \frac{\pi}{8}\right)}}, & \text{if } (f,\theta,\varphi)\in \left[-\frac W2,0\right]\\[-5pt]
 & \times \left[0, \frac \pi4\right]\times \left[-\frac \pi4,0\right]\\
 & \text{or if } (f,\theta,\varphi)\in \left[-\frac W2, \frac W2\right]\\
 & \times \left[-\frac\pi4, -\frac\pi8\right] \times \left[\frac\pi8,\frac\pi4\right]\\
 0, & \text{otherwise.}
\end{cases} \label{desired_bp}
\end{equation}
Hence, the power radiated with the desired beampattern is
\begin{equation}
 \frac{1}{4\pi} \int_{-W/2}^{W/2} \int_{-\pi/2}^{\pi/2}\int_{-\pi/2}^{\pi/2}  D^2(f;\theta, \varphi) \cos \theta d\theta d\varphi df  = \SI{8}{\W}. \label{radiated_power_d}
\end{equation}
The available power is $P=\SI{10}{\W}$, and the weighting function is $\tilde w(f;\theta,\varphi)=1$, so that, from~\eqref{radiated_power_d}, Eq.~\eqref{weights_expr} becomes $w(f;\theta,\varphi)=(W\pi \cos\theta) /(32KL)$. The number of frequency sampling points is $K=128$, and they are uniformly spaced in\footnote{Observe that 128 is roughly twice the number of sampling points needed to (uniformly) sample the source signals at the Nyquist rate.} $[-W/2, W/2]$; in the angular domain, $36 \times 36$ sampling points uniformly spaced in $[-\pi/2, \pi/2]^2$ are taken,\footnote{Notice that the smallest half-power beamwidth of a uniformly excited, equally spaced linear array with 10 elements and half-wavelength spacing is approximately equal to $0.891 \frac{2}{10}\approx \SI{0.1782}{\radian}$~\cite[Eq.~(2.100)]{Van_Trees_4}; therefore, 36 is roughly twice the number of sampling points needed to uniformly sample the interval $[-\pi/2,\pi/2]$ with a \SI{0.1782}{\radian} spacing.} so that $L=1296$. 

In all figures, we plot the normalized power beampattern (NPB), i.e., the power beampattern normalized by its maximum value
\begin{equation}
  \text{NPB}(f;\theta,\varphi) = \frac{B^2(f;\theta,\varphi)}{\max\limits_{(f',\theta',\varphi')} B^2(f';\theta',\varphi')}.
\end{equation}
Additional performance metrics used to assess the performance of the synthesized beampattern are the integrated sidelobe level (ISL) and the peak sidelobe level (PSL):
\begin{subequations}
\begin{align}
 \text{ISL}&= \frac{\frac{1}{4\pi} \iiint\limits_{\mathcal S} B^2(f;\theta,\varphi) \cos \theta d\theta d\varphi df}{\frac{1}{4\pi} \iiint\limits_{\mathcal M} B^2(f;\theta,\varphi) \cos \theta d\theta d\varphi df},\\
 \text{PSL} &= \frac{\max\limits_{(f,\theta,\varphi)\in \mathcal S} B^2(f;\theta,\varphi)}{\max\limits_{(f,\theta,\varphi)} B^2(f;\theta,\varphi)},
\end{align}%
\end{subequations}
where $\mathcal M$ is the region where the desired beampattern in~\eqref{desired_bp} is non-zero (we can call it mainlobe), and $\mathcal S$ is the sidelobe region, defined here as the complement of the mainlobe region after it has been enlarged with a guard interval of $\pi/16$ for the azimuth and elevation and 1~MHz for the frequency). Finally, the convergence criterion adopted for Alg.~\ref{block-coordinate_descent_alg} is to stop when either the objective function or all the dependent variables experience a relative change between consecutive iterations smaller than $10^{-6}$.

\subsection{Benchmark}\label{sec_competitive_systems}

For the sake of comparison, we also consider the cases of a fully-digital MIMO system, a hybrid analog-digital MIMO system with 4 sources, and a phased array (PA), all having $M$ elements with position $\bm p_i$ and power beampattern $4\cos \theta \cos \varphi$, $i=1,\ldots,M$. Let $\zeta_i(t)$ be the waveform sent over the $i$-th element of the array, and define
\begin{subequations}
 \begin{align}
  \bm \zeta_n &= \begin{bmatrix} \zeta_1 (n/W) & \cdots & \zeta_M(n/W) \end{bmatrix}\transp \in \mathbb C^M, \\
  \bm \zeta &=\begin{bmatrix}\bm \zeta_1\transp & \cdots & \bm \zeta_N\transp \end{bmatrix}\transp \in \mathbb C^{MN}, \\
   \tilde{\bm Q} (f; \theta, \varphi)& = \frac{\sqrt{4\cos \theta \cos \varphi}}{W \sqrt T} \bigl( \bm e\transp(f) \otimes \bm I_M \bigr) \in \mathbb C^{M\times MN}.
 \end{align}%
\end{subequations}
Then, it is not difficult to show that the amplitude beampattern of the MIMO system and of the PA can be expressed in both cases as $| \bm v\herm(f; \theta, \varphi) \tilde{\bm Q}(f; \theta, \varphi) \bm \zeta |$. Such a beampattern is optimized according to the same criterion used for the illuminator and RIS architecture considered in Sec.~\ref{sys_model_sec}.

For the MIMO system, the optimization problem is
\begin{equation}
 \begin{aligned}
 \min_{\bm \zeta \in \mathbb C^{MN}} & \; \sum_{k=1}^K \sum_{\ell=1}^L w_{k\ell} \bigl( D_{k\ell} - | \bm v_{k\ell}\herm \tilde{\bm Q}_{k\ell} \bm \zeta | \bigr)^2,\\
 \text{s.t.} & \; \Vert \bm \zeta \Vert^2 \leq N P,
 \end{aligned} \label{opt_prob_MIMO}
\end{equation}
where $\tilde{\bm Q}_{k\ell} = \tilde{\bm Q}(f_k;\theta_\ell, \varphi_\ell)$. If constant-modulus waveforms are required, we have that $\bm \zeta = (\bm 1_N \otimes \bm \gamma) \odot \bm \xi$, where $\bm \gamma \in\mathbb R^M$ is the amplitude vector over the $M$ elements, and $\bm \xi \in \mathbb C^{MN}$ is the vector containing the unit-modulus complex scalars accounting for the phase terms, and the problem is
\begin{equation}
 \begin{aligned}
  \min_{\substack{\bm \gamma\in\mathbb R^M, \bm \xi \in \mathbb C^{MN}, \\\bm \zeta \in \mathbb C^{MN}}} & \; \sum_{k=1}^K \sum_{\ell=1}^L w_{k\ell} \bigl( D_{k\ell} -  | \bm v_{k\ell}\herm \tilde{\bm Q}_{k\ell} \bm \zeta | \bigr)^2,\\
 \text{s.t.} & \; \bm \zeta = (\bm 1_N \otimes \bm \gamma) \odot \bm \xi,\\
  & \; \Vert \bm \gamma \Vert^2 \leq P,\\
  & \; | \xi_i | = 1, \quad i=1,\ldots,MN.
 \end{aligned} \label{opt_prob_MIMO_const_mod}
\end{equation}
These problems are solved as discussed in Secs.~\ref{min_prob_general} and~\ref{constr_s_sec}.

For the PA, the space-time signal $\bm \zeta$ is bounded to have a special structure, i.e., $\bm \zeta = \bm \varrho \otimes \bm \chi$, where $\bm \varrho \in \mathbb C^N$ represents the transmit waveform, and $\bm \chi \in\mathbb C^M$, with $|\chi_i|=1$ for all $i$, is the beamforming vector. Accordingly, the corresponding optimization problems are
\begin{equation}
 \begin{aligned}
 \min_{\substack{\bm \varrho \in \mathbb C^N , \bm \chi \in\mathbb C^M \\ \bm \zeta \in \mathbb C^{MN}}} & \; \sum_{k=1}^K \sum_{\ell=1}^L w_{k\ell} \bigl( D_{k\ell} - | \bm v_{k\ell}\herm \tilde{\bm Q}_{k\ell} \bm \zeta | \bigr)^2,\\
 \text{s.t.} & \; \bm \zeta = \bm \varrho \otimes \bm \chi,\\
 & \; |\chi_i|=1, \quad i=1,\ldots, M,\\
 & \; \Vert \bm \varrho \Vert^2 \leq NP/M,
 \end{aligned} \label{opt_prob_PA}
\end{equation}
and, if constant-modulus waveforms are required,
\begin{equation}
 \begin{aligned}
 \min_{\substack{\bm \varrho \in \mathbb C^N , \bm \chi \in\mathbb C^M \\ \alpha \in \mathbb R, \bm \zeta \in \mathbb C^{MN}}} & \; \sum_{k=1}^K \sum_{\ell=1}^L w_{k\ell} \bigl( D_{k\ell} -  | \bm v_{k\ell}\herm \tilde{\bm Q}_{k\ell} \bm \zeta | \bigr)^2,\\
 \text{s.t.} & \; \bm \zeta = \alpha \bm \varrho \otimes \bm \chi,\\
 & \; |\chi_i|=1, \quad i=1,\ldots, M,\\
 & \; |\varrho_i|=1, \quad i=1,\ldots, N,\\
 & \; \alpha^2 \leq P/M.
 \end{aligned} \label{opt_prob_PA_const_mod}
\end{equation}
These problems are again solved as in Secs.~\ref{min_prob_general} and~\ref{constr_s_sec}, where, for Problem~\eqref{opt_prob_PA_const_mod}, the minimization over the coordinate variable $\alpha$ simply gives
\begin{multline}
 \alpha = \min \Biggl\{\sqrt{\frac{P}{M}}, \\\frac{\sum_{k=1}^K \sum_{\ell=1}^L w_{k\ell} D_{k\ell}\Re(\e^{-\i\psi_{k\ell}} \bm v_{k\ell}\herm \tilde{\bm Q}_{k\ell} (\bm  \varrho \otimes \bm \chi))}{\sum_{k=1}^K \sum_{\ell=1}^L w_{k\ell} |\bm v_{k\ell}\herm \tilde{\bm Q}_{k\ell} (\bm \varrho \otimes \bm \chi) |^2}\Biggr\}.
\end{multline}

Finally, as for the hybrid MIMO system with four sources, the array is partitioned into four subarrays with an equal number of antennas. Each subarray contains the antennas from one of the four quadrants and is connected to one source. Then, it is not difficult to show that the amplitude beampattern of the hybrid MIMO is the same as the RIS-based architecture once we redefine the channel frequency responses in~\eqref{channel_freq_resp} as $G_{ij}(f)=1$, if source $j$ and antenna $i$ belong to the same quadrant, and $G_{ij}(f)=0$, otherwise. The optimization of this system is therefore carried out as in the RIS-based architecture.

\subsection{Computational complexity analysis}\label{complexity_Sec}
The RIS-based architecture is optimized by solving Problem~\eqref{problem_2}; a suboptimum solution is described in Sec.~\ref{min_prob_general}, and the complete routine is reported in Alg.~\ref{block-coordinate_descent_alg}. Here, we evaluate the computational complexity per iteration of this routine. Specifically, the cost for evaluating $\bm A$ and $\bm b$ in~\eqref{def_A_b} is $\mathcal O \bigl(KL(MJN +J^2N^2) \bigr)$; the cost for solving Problem~\eqref{sub_prob_s_2} is dominated by the eigenvalue decomposition and is $\mathcal O (J^3N^3)$; the cost for evaluating $\bm B$ and $\bm c$ in~\eqref{def_B_c} is $\mathcal O \bigl(KL (MJN + M^2)\bigr)$; the cost for updating $\{x_i\}_{i=1}^M$ is $\mathcal O (M^2)$; and the cost for updating $\{\psi_{k\ell}\}_{k,\ell}$ is $\mathcal O (KLMJN)$. Therefore, the overall computational complexity of Alg.~\ref{block-coordinate_descent_alg} per iteration is $\mathcal O\bigl(KL(MJN +J^2N^2+M^2) +J^3N^3\bigr)$.

If constant-modulus signals are required, the RIS-based architecture is optimized by solving Problem~\eqref{problem_constant_mod}. As described in Sec.~\ref{constr_s_sec}, a suboptimum solution can be found by exploiting the block-coordinate ascent algorithm in Sec.~\ref{min_prob_general} with appropriate block-variables. In this case, the computational complexity per iteration is $\mathcal O\bigl(KL(MJN +J^2N^2+M^2) +J^3\bigr)$.

Regarding the systems considered for comparison, the MIMO system is optimized by solving Problem~\eqref{opt_prob_MIMO}, and Problem~\eqref{opt_prob_MIMO_const_mod} if constant-modulus signals are required. If these problems are solved as in Secs.~\ref{min_prob_general} and~\ref{constr_s_sec}, the computational complexity per iteration is $\mathcal O (KLM^2N^2+M^3N^3)$ for Problem~\eqref{opt_prob_MIMO}, and $\mathcal O (KLM^2N^2+M^3)$ for Problem~\eqref{opt_prob_MIMO_const_mod}. The PA is optimized by solving Problem~\eqref{opt_prob_PA}, and Problem~\eqref{opt_prob_PA_const_mod} if constant-modulus signals are required. Using the approach described in Secs.~\ref{min_prob_general} and~\ref{constr_s_sec}, the computational complexity per iteration is $\mathcal O (KLM^2N+N^3)$ per iteration for Problem~\eqref{opt_prob_PA}, and $\mathcal O (KLM^2N+N^2)$ for Problem~\eqref{opt_prob_PA_const_mod}. Finally, the computational complexity of the hybrid MIMO system is the same as that of the RIS-based one.

\begin{table}[t]{
\caption{Computational complexity of the approach in Sec.~\ref{min_prob_general}}\label{tab_complexity}
\begin{center}
\begin{tabular}{llll}
  \toprule & System & Problem & Complexity per iteration\\
  \midrule & \cc RIS-based & \cc  & \cc $\mathcal O\bigl(KL(MJN +J^2N^2$ \\
  & \cc \& Hybrid MIMO & \cc \multirow{-2}{*}{\eqref{problem_2}}& \cc $\hphantom{\mathcal O\bigl(} +M^2) +J^3N^3\bigr)$\\
  & MIMO & \eqref{opt_prob_MIMO} & $\mathcal O (KLM^2N^2+M^3N^3)$ \\
  & \cc PA & \cc \eqref{opt_prob_PA} & \cc$\mathcal O (KLM^2N+N^3)$ \\
  \midrule \multirow{3}[3]{3em}{Const. \\mod. \\ signals} & RIS-based & & $\mathcal O\bigl(KL(MJN +J^2N^2$\\
  & \& Hybrid MIMO & \multirow{-2}{*}{\eqref{problem_constant_mod}} & $\hphantom{\mathcal O\bigl(}+M^2) +J^3\bigr)$ \\
  & \cc MIMO & \cc \eqref{opt_prob_MIMO_const_mod} & \cc $\mathcal O (KLM^2N^2+M^3)$\\
  & PA & \eqref{opt_prob_PA_const_mod} & $\mathcal O (KLM^2N+N^2)$\\
  \bottomrule
 \end{tabular}
\end{center}}
\end{table}

To facilitate the comparison, Table~\ref{tab_complexity} summarizes the computational complexity of all analyzed systems.

\subsection{Numerical results}\label{num_res_sec}

\begin{figure}[t]	
\centerline{\includegraphics[width=\columnwidth]{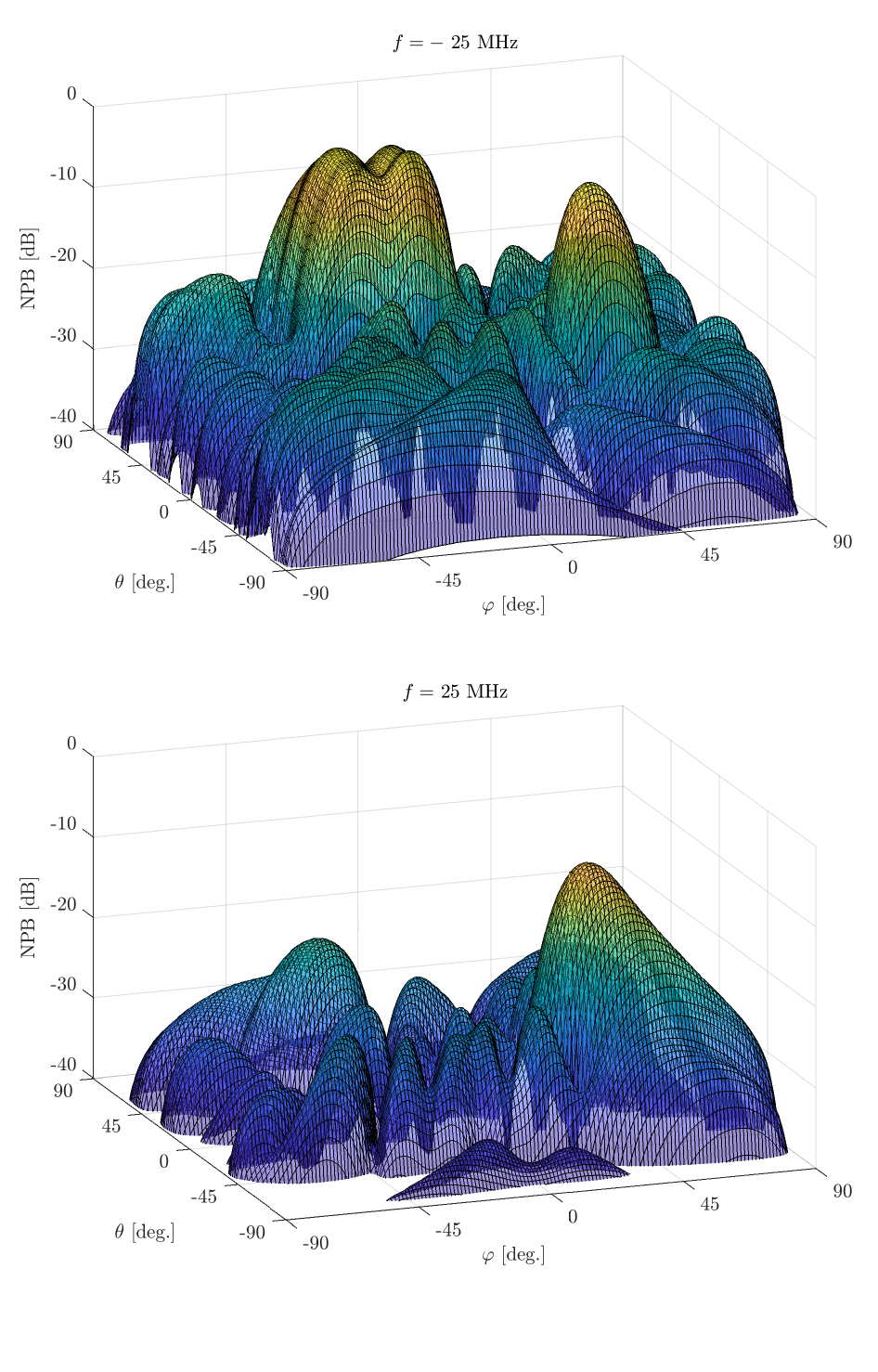}}	
 \caption{Normalized power beampattern (synthesized with the RIS-based architecture) as a function of elevation and azimuth at frequencies $-25$ MHz (top) and 25 MHz (bottom).} \label{fig_2}
\end{figure}

The following figures have been drawn using 128 frequency grid points uniformly spaced in $[-W/2, W/2]$ and $144 \times 144$ grid points uniformly spaced in $[-\pi, \pi]^2$. Fig.~\ref{fig_2} shows a 3-D plot of the NPB of the RIS-based architecture when no unit-modulus constraint is imposed on the illuminator signals. The wide beam for the negative frequencies and the narrow beam for all frequencies are clearly visible and well-shaped, and the sidelobes are relatively low.

\begin{figure}[t]	
\centerline{\includegraphics[width=\columnwidth]{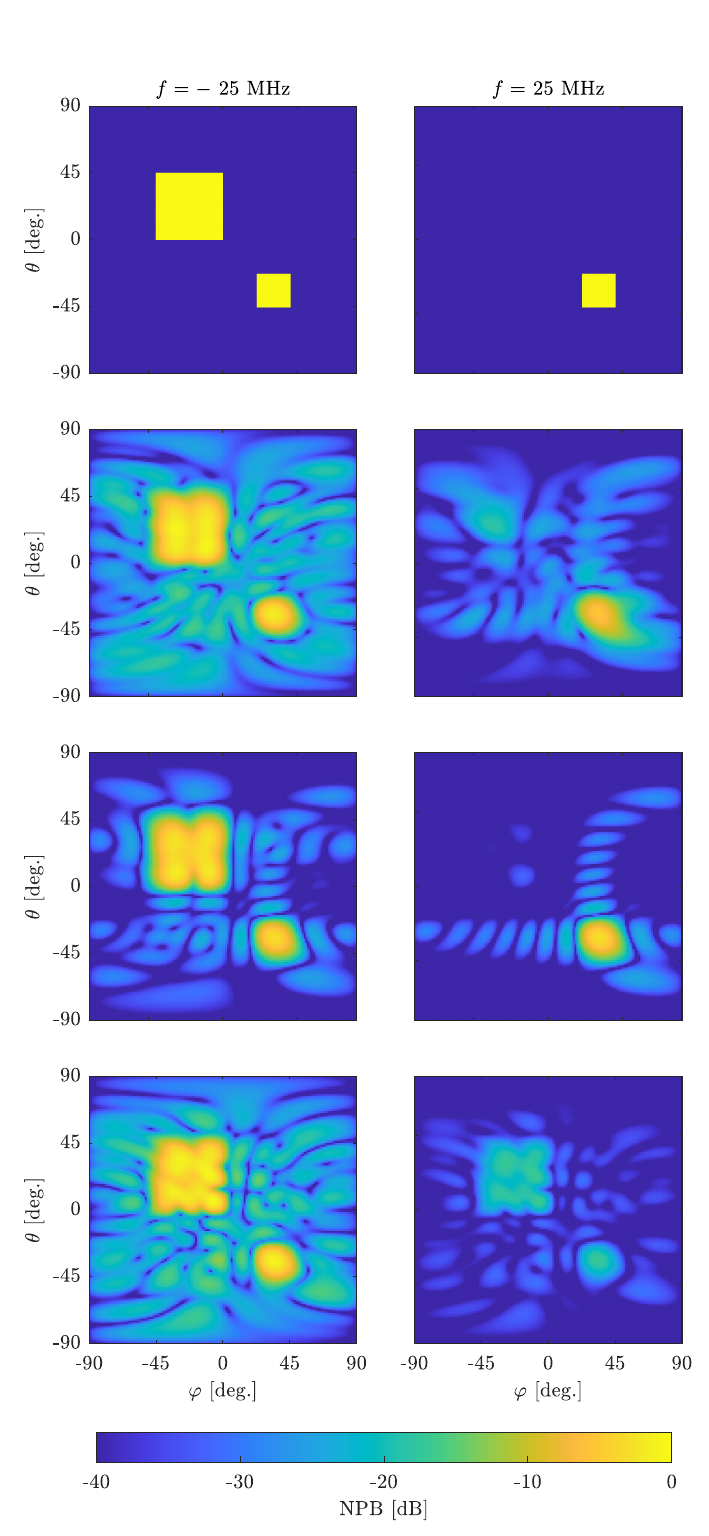}}	
 \caption{Desired (top row) and synthesized with the RIS-based architecture (second row), with a MIMO system (third row), and with a phased array (bottom row) normalized power beampattern as a function of elevation and azimuth at frequencies $-25$ MHz (left column) and 25 MHz (right column).} \label{fig_3}
\end{figure}

\begin{figure}[t]	
\centerline{\includegraphics[width=\columnwidth]{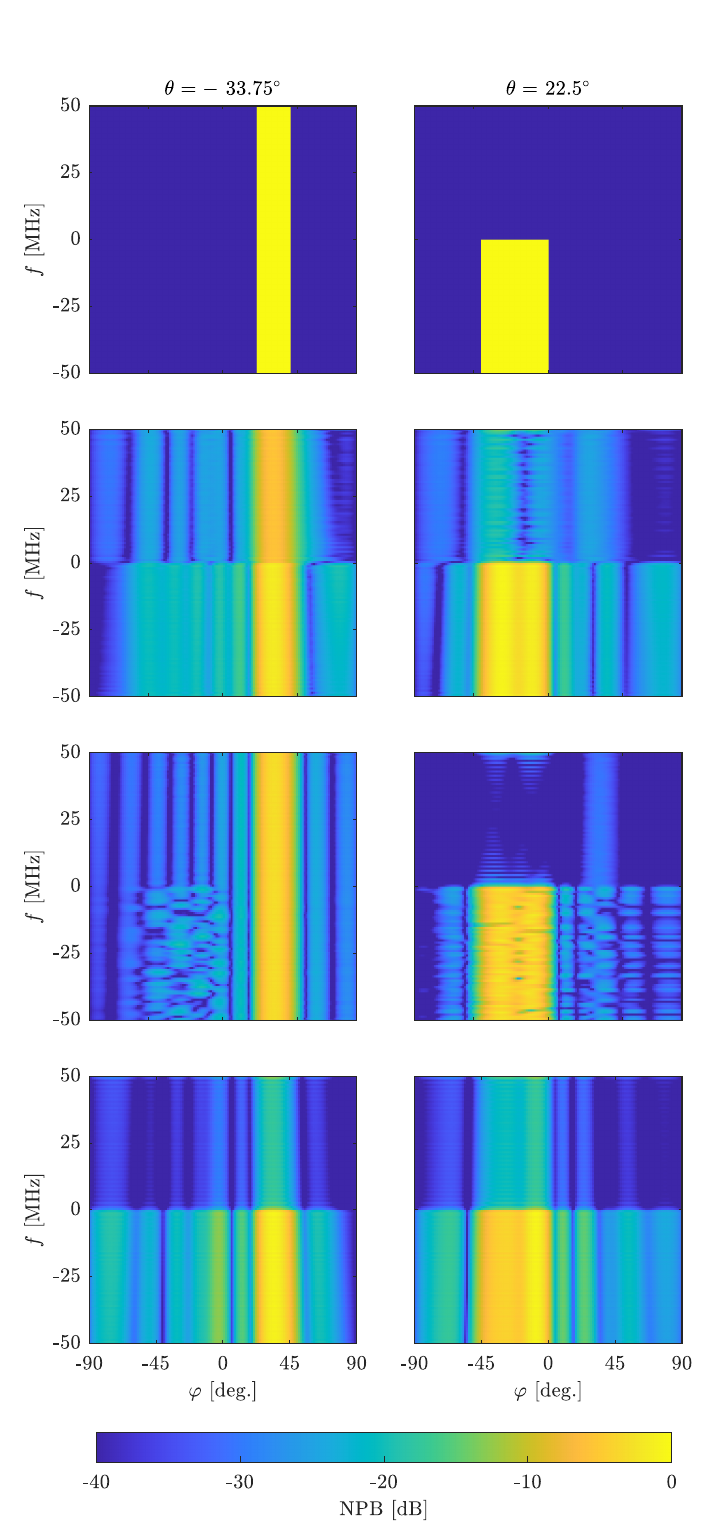}}	
 \caption{Desired (top row) and synthesized with the RIS-based architecture (second row), with a MIMO system (third row), and with a phased array (bottom row) normalized power beampattern as a function of frequency and azimuth at elevations $\SI{-33.75}{\degree}$ (left column) and $\SI{22.5}{\degree}$ (right column).} \label{fig_4}
\end{figure}

\begin{figure}[t]	
\centerline{\includegraphics[width=\columnwidth]{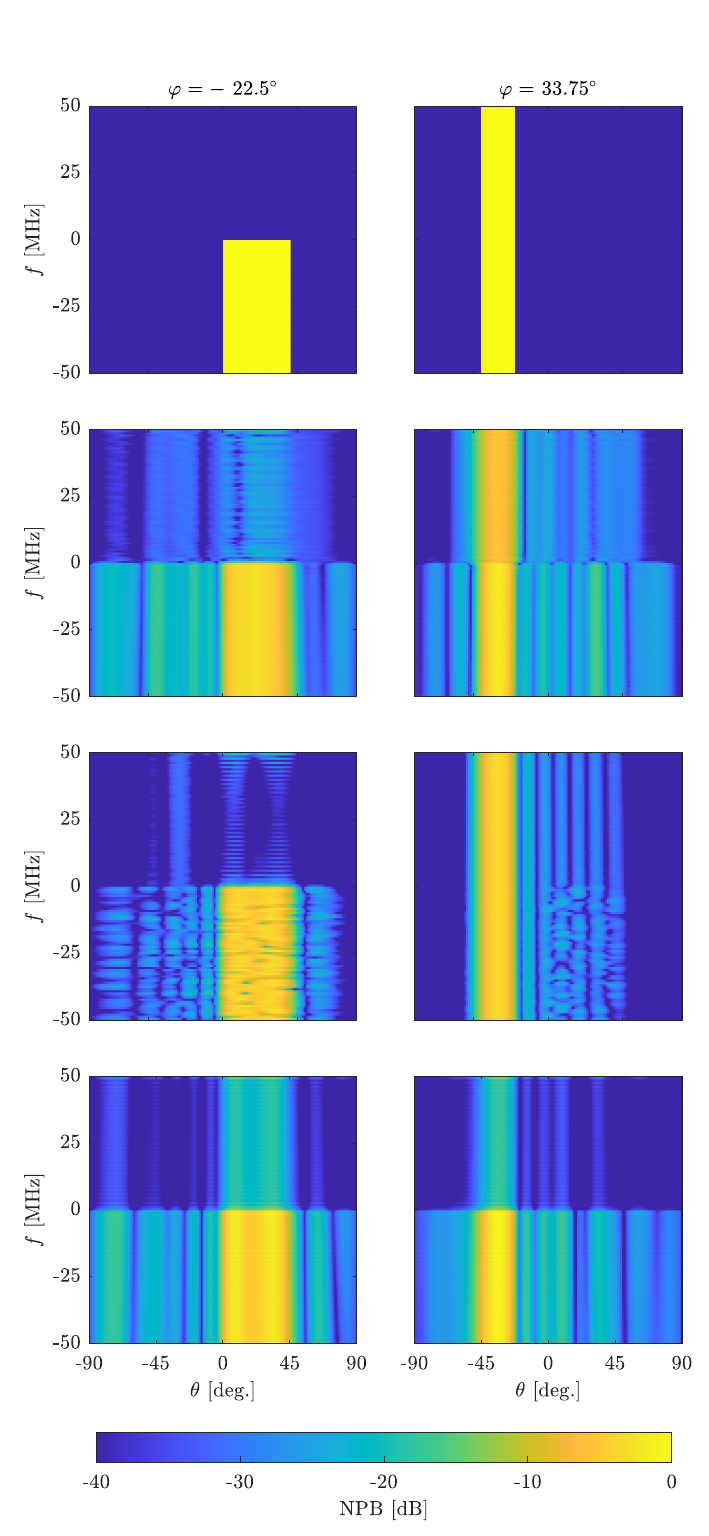}}	
 \caption{Desired (top row) and synthesized with the RIS-based architecture (second row), with a MIMO system (third row), and with a phased array (bottom row) normalized power beampattern as a function of frequency and elevation at azimuths $\SI{-22.5}{\degree}$ (left column) and $\SI{33.75}{\degree}$ (right column).} \label{fig_5}
\end{figure}

The next figures report 2-D intensity plots of the desired NPB (first row) and of the NPB synthesized with the RIS-based transmit architecture (second row) and with the MIMO system (third row) and PA (fourth row) benchmarks. In all cases, no unit-modulus constraint is imposed on the illuminator signals. In particular, Figs.~\ref{fig_3},~\ref{fig_4}, and~\ref{fig_5} show the NPB as a function of the azimuth and elevation for two frequencies ($\SI{-25}{\MHz}$ in the left column, $\SI{25}{\MHz}$ in the right column), as a function of the frequency and azimuth for two elevations ($\SI{-33.75}{\degree}$ in the left column, $\SI{22.5}{\degree}$ in the right column), and as a function of the frequency and elevation for two azimuths ($\SI{-22.5}{\degree}$ in the left column, $\SI{33.75}{\degree}$ in the right column), respectively. Thanks to the higher number of degrees of freedom, the MIMO system is able to realize the beampattern most similar to the desired one, with well-shaped beams, sharp transitions between the two frequency regions, and low sidelobes. The worst performance is obtained with the PA, due to the limited number of degrees of freedom for the beampattern design problem, and this is particularly evident in the frequency domain.\footnote{Here, we are only concerned with the mathematical model, and PA is just a particular case of MIMO, so its performance is necessarily worse due to the reduced number of degrees of freedom. However, PAs are not outdated (especially in radar applications) due to their simplicity and reliability in challenging environments.} The results achieved with the RIS-based architecture lie between these two cases.

\begin{figure}[t]	
\centerline{\includegraphics[width=\columnwidth]{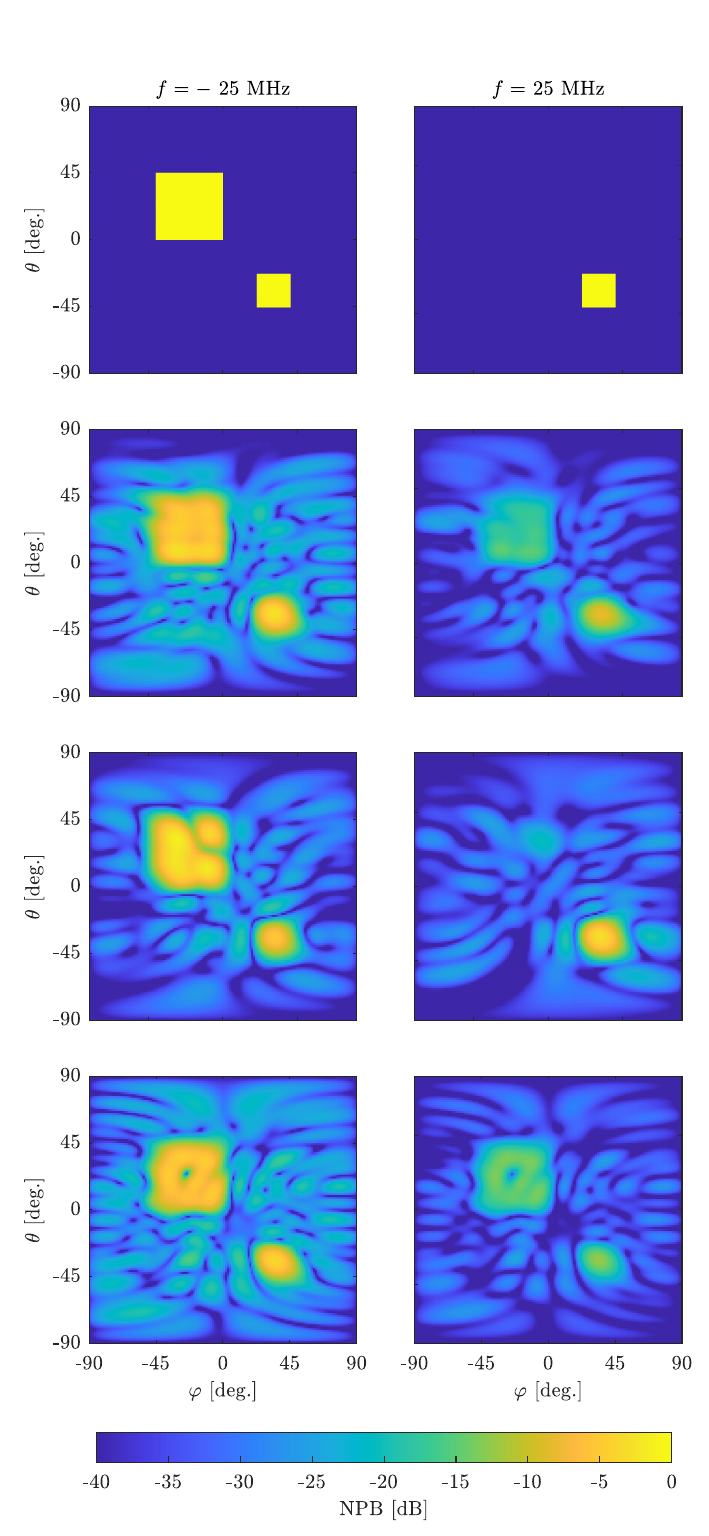}}	
 \caption{Desired (top row) and synthesized with the RIS-based architecture (second row), with a MIMO system (third row), and with a phased array (bottom row) normalized power beampattern as a function of elevation and azimuth at frequencies $-25$ MHz (left column) and 25 MHz (right column) when constant-modulus illuminator signals are used.} \label{fig_6}
\end{figure}

\begin{figure}[t]	
\centerline{\includegraphics[width=\columnwidth]{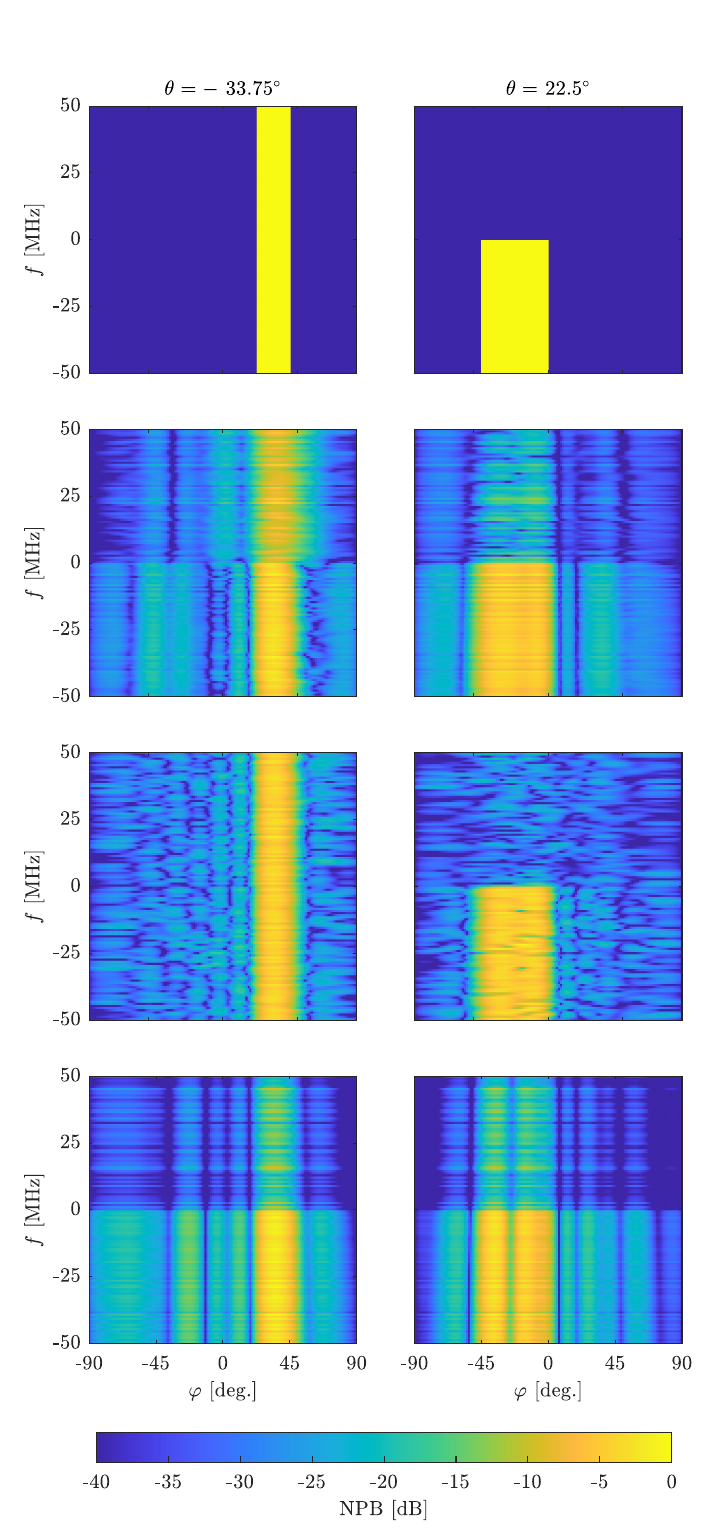}}	
 \caption{Desired (top row) and synthesized with the RIS-based architecture (second row), with a MIMO system (third row), and with a phased array (bottom row) normalized power beampattern as a function of frequency and azimuth at elevations $\SI{-33.75}{\degree}$ (left column) and $\SI{33.75}{\degree}$ (right column) when constant-modulus illuminator signals are used.} \label{fig_7}
\end{figure}

\begin{figure}[t]	
\centerline{\includegraphics[width=\columnwidth]{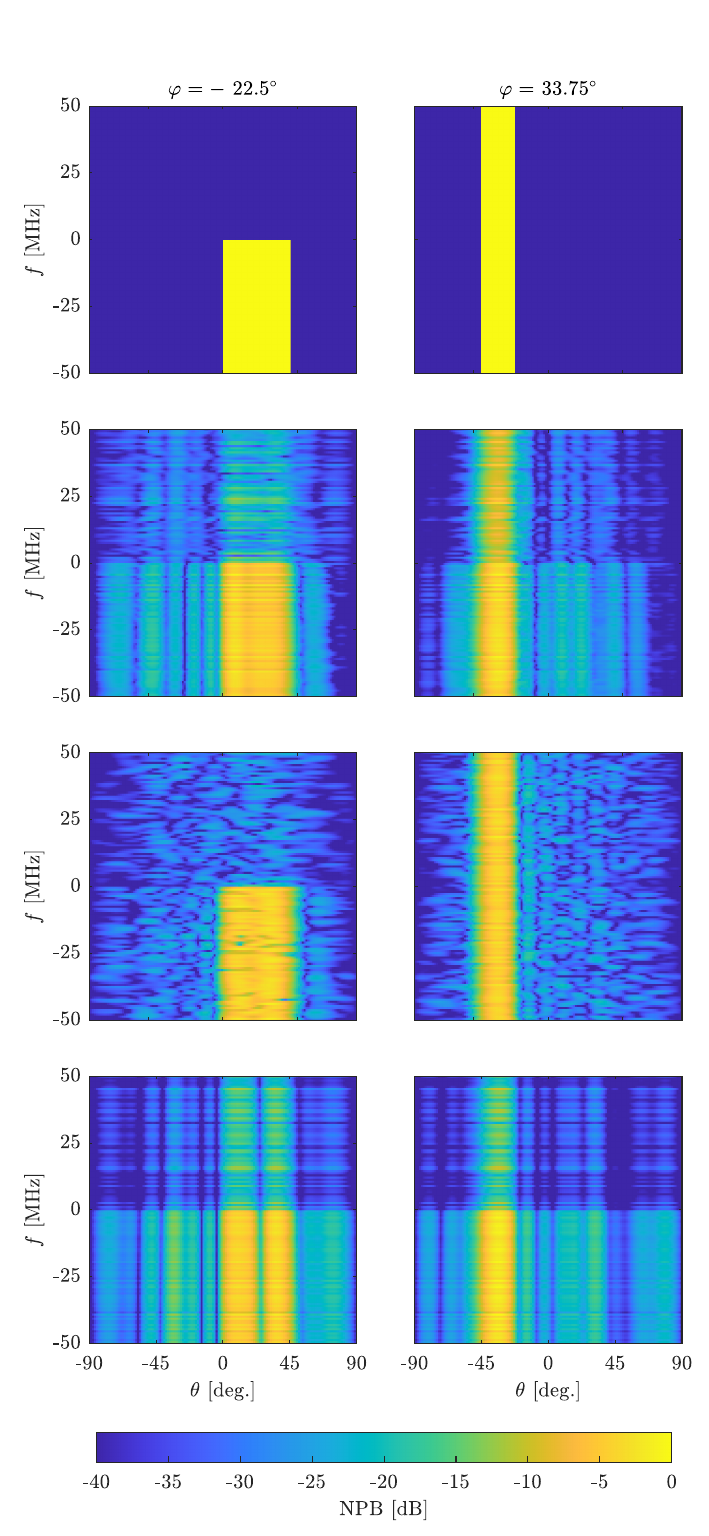}}	
 \caption{Desired (top row) and synthesized with the RIS-based architecture (second row), with a MIMO system (third row), and with a phased array (bottom row) normalized power beampattern as a function of frequency and elevation at azimuths $\SI{-33.75}{\degree}$ (left column) and $\SI{33.75}{\degree}$ (right column) when constant-modulus illuminator signals are used.} \label{fig_8}
\end{figure}

Figs.~\ref{fig_6},~\ref{fig_7}, and~\ref{fig_8} replicate Figs.~\ref{fig_3},~\ref{fig_4}, and~\ref{fig_5} for the case where constant-modulus illuminator signals are used. Due to the additional constraint, the realized beampattern is worse for all three systems, with a higher ripple in the frequency domain. Specifically, the MIMO system exhibits the largest relative loss compared to the corresponding non-constant-modulus case, as can also be seen from Table~\ref{tab_RSE}. The RIS-based architecture, which already has an \emph{intrinsic} constant-modulus constraint on the RIS phases, better bears the reduction in the available degrees of freedom caused by the additional constant-modulus constraint of the source signals. In particular, the beampattern realized with the RIS-based architecture using only 4 active sources does not look very dissimilar to the one realized with the MIMO system that exploits 100 active sources, as can also be seen from the RSE.

Table~\ref{tab_RSE} shows the RSE of all considered systems (the RIS-based architecture and the MIMO, hybrid MIMO, and PA benchmarks), with and without the constant-modulus constraint on the illuminator signals, for different numbers of sources (1 and 4) and different numbers of RIS elements (radiating antennas for the benchmarks).\footnote{The power radiated by the sources (not reported in the table) to obtain the desired beampattern is approximately equal to the available power for all considered systems. Notice, however, that the radiated power is, in general, a small fraction of the overall power budget needed to operate the arrays, and fully digital MIMO systems are known to be much more demanding in terms of overall power requirement than RIS-based transmitters~\cite{Jamali_2021}.} Notice that the single-source RIS-based transmit architecture mirrors the system analyzed in~\cite{Rahal_2022} for the narrowband scenario without waveform optimization. The performance monotonically increases with $M$ for all systems and with $J$ for the RIS-based architecture. As qualitatively evaluated in Figs.~\ref{fig_5}--\ref{fig_8}, the quantitative analysis in Table~\ref{tab_RSE} confirms that the MIMO system achieves the highest fidelity to the desired beampattern, and that the RIS-based architecture lies between the MIMO system and the PA. Interestingly, the four-sources RIS-based architecture outperforms the four-sources hybrid MIMO system. This seemingly unexpected outcome can be attributed to the system's capability to generate complex amplitude distributions across the surface elements through the carefully controlled constructive and destructive interference of the signals emitted by the four sources. This fine-grained control allows for more effective beam-shaping than the hybrid MIMO system, which cannot independently adjust the amplitude of signals directed to individual antenna elements within a quadrant. Furthermore, the RSE of the 4-sources RIS-based architecture with $M=60$ and $M=100$ elements is smaller than the RSE of the MIMO system with $M=36$ active antennas, which requires a number of (costly) RF-chains nine times larger.

\begin{table}[t]{
\caption{Relative square error: All systems}\label{tab_RSE}
\begin{center}
\begin{tabular}{llccc}
  \toprule & System & $M=36$ & $M=64$ & $M=100$\\
  \midrule & \cc RIS-based ($J=1$) & \cc 0.356 & \cc 0.275 & \cc 0.238\\
  & RIS-based ($J=4$) & 0.305 & 0.241 & 0.208\\
  & \cc MIMO & \cc 0.249 & \cc 0.179 & \cc 0.133\\
  & Hybrid MIMO ($J=4$) & 0.334 & 0.249 & 0.239 \\
  & \cc PA & \cc 0.371 & \cc 0.297 & \cc 0.270\\
  \midrule & RIS-based ($J=1$) & 0.368 & 0.292 & 0.254 \\
  Const. & \cc RIS-based ($J=4$) & \cc 0.317 & \cc 0.261 & \cc 0.237\\
  mod. & MIMO & 0.272 & 0.223 & 0.179\\
  signals & \cc Hybrid MIMO ($J=4$) & \cc 0.354 & \cc 0.272 & \cc 0.266 \\
  & PA & 0.415 & 0.342 & 0.293\\
  \bottomrule
 \end{tabular}
\end{center}}
\end{table}

Finally, Table~\ref{tab_sidelobes} presents the PSL and ISL of the RIS-based transmit architecture for different numbers of sources ($J = 1,4$) and RIS elements ($M = 36, 64, 100$), with and without a constant-modulus constraint on the source signals. It can be seen that the performance improves with $M$ and $J$; however, the improvement observed in PSL and ISL across the tested configurations is less substantial compared to the variation seen in the RSE, which is the performance metric optimized in this work.

\section{Conclusion}\label{conclusion_sec}

In this work, we have investigated the problem of beampattern design in a transmit system leveraging a large reflecting/transmitting reconfigurable intelligent surface (RIS) driven by a small number of active sources. Our analysis demonstrates that this RIS-based architecture can achieve satisfactory performance, exhibiting a low relative square error between the desired and realized beampatterns. Notably, our simulation results with a 4-source, 100-element RIS reveal that the generated beampattern is not dramatically different from that of a conventional MIMO system with 100 active elements and is better than that of a 36-element MIMO system. This highlights the significant potential of the RIS-based architecture for cost reduction in applications requiring sophisticated beampattern synthesis. Future research should focus on optimizing the number and placement of the sources, exploring the benefits of active RISs (which offer enhanced performance, especially in scenarios with substantial source-RIS separation), and investigating system design for near-field communication scenarios~\cite{Chen_2024}. Furthermore, a key direction should be the development and thorough experimentation with a physical prototype~\cite{Pei_2021, Fara_2022, Peng_2025, Tishchenko_2025} to rigorously validate the theoretical findings presented in this study.

\begin{table}[t]{
\caption{Peak and Integrated sidelobe levels [dB]:\\RIS-based architecture} \label{tab_sidelobes}
\begin{center}
\begin{tabular}{lccccc}
  \toprule & & \multicolumn{2}{c}{$J=1$} & \multicolumn{2}{c}{$J=4$}  \\
  \cmidrule(lr){3-4} \cmidrule(lr){5-6} & $M$ & PSL & ISL & PSL & ISL \\
  \midrule & \cc 36  & \cc $-12.1$ & \cc $-9.46$ & \cc $-12.6$ & \cc $-10.1$\\
  & 64 & $-12.6$ & $-11.0$ & $-13.0$ & $-11.7$\\
  & \cc 100 & \cc $-12.8$ & \cc $-11.7$ & \cc $-13.1$ & \cc $-11.8$\\
  \midrule  Const. & 36 & $-11.5$ & $-8.64$ & $-12.1$ & $-10.0$\\
  mod. & \cc 64 & \cc $-11.7$ & \cc $-9.80$ & \cc $-12.1$ & \cc $-10.0$\\
  signals & 100 & $-11.8$ & $-10.3$ & $-12.3$ & $-10.1$\\
  \bottomrule
 \end{tabular}
\end{center}}
\end{table}

\appendix\label{appendix}

Problem~\eqref{sub_prob_s_2} is the minimization of a convex quadratic function ($\bm A$ is a Hermitian positive semidefinite matrix) over a ball, and its solution in terms of a diagonalization of $\bm A$ is standard. Indeed, recalling that $\tilde{\bm s}= \bm U\herm \bm s$ and $\tilde{\bm b}= \bm U\herm \bm b$, Problem~\eqref{sub_prob_s_2} can be recast as
\begin{equation}
 \begin{aligned}
 \min_{\tilde{\bm s} \in \mathbb C^{JN}} & \; \sum_{i=1}^{JN} \left( \sigma_i |\tilde s_i|^2 -2 \Re( \tilde b_i \tilde s_i^*)\right), \\
 \text{s.t.} & \; \sum_{i=1}^{JN} |\tilde s_i|^2 \leq NP,
 \end{aligned}
\end{equation}
and the Karun-Kush-Tucker conditions are
\begin{equation}
 \begin{cases}
  (\sigma_i + \lambda ) \tilde s_i = \tilde b_i, \quad i=1,\ldots,JN,\\
  \lambda \geq 0,\\
  \sum_{i=1}^{JN} |\tilde s_i|^2 \leq NP,\\
  \lambda \left(\sum_{i=1}^{JN} |\tilde s_i|^2 - NP\right)=0.
 \end{cases}
\end{equation}

At this point, upon introducing the function
\begin{equation}
 f(\lambda) = \sum_{i=1}^{JN} \frac{|\tilde b_i|^2}{(\sigma_i+\lambda)^2},
\end{equation}
we can distinguish between the following cases.
\begin{enumerate}
 \item If $\sigma_i>0$ for all $i$ and $f(0) \leq NP$, we can take $\lambda =0$ and then $\tilde s_i= \tilde b_i/\sigma_i$;
 
 \item If $\sigma_i>0$ for all $i$ and $f(0) > NP$, then $\lambda>0$, and it must be chosen so that $f(\lambda) = NP$, while $\tilde s_i = \tilde b_i/ (\sigma_i + \lambda)$, for any $i$; notice that, since $f$ is strictly decreasing, and $f(0)>N P$, the equation $f(\lambda) = NP$ admits a unique solution.
 
 \item If $\sigma_i = 0$ for some $i$, and there exists $i$ such that $\sigma_i=0$ and $\tilde b_i\neq 0$, then $\lambda>0$, and it must be chosen so that $f(0) = NP$, while $\tilde s_i = \tilde b_i/ (\sigma_i + \lambda)$; since $f$ is strictly decreasing, and $\lim_{\lambda\rightarrow 0} f(\lambda)=\infty$, the equation $f(\lambda) = NP$ admits a unique solution.
 
 \item If $\sigma_i = 0$ for some $i$, $\tilde b_i= 0$ for all $i$ such that $\sigma_i=0$, and $\sum_{i=1,i:\sigma_i>0}^{JN} |\tilde b_i|^2/\sigma_i^2 \leq NP$, we can take $\lambda =0$ and then $\tilde s_i= \tilde b_i/\sigma_i$, if $\sigma_i>0$, while it can be arbitrarily set equal to zero, if $\sigma_i=0$.
 
 \item If $\sigma_i = 0$ for some $i$, $\tilde b_i= 0$ for all $i$ such that $\sigma_i=0$, and $\sum_{i=1,i:\sigma_i>0}^{JN} |\tilde b_i|^2/\sigma_i^2 > NP$, then $\lambda>0$, and it must be chosen so that $f(0) = NP$, while $\tilde s_i = \tilde b_i/ (\sigma_i + \lambda)$; since $f$ is strictly decreasing, and $\lim_{\lambda\rightarrow 0} f(\lambda)=\infty$, the equation $f(\lambda) = NP$ admits a unique solution.
\end{enumerate}
Finally, these five cases can be condensed as follows.
\begin{itemize}
 \item If $\sum_{i=1, \sigma_i\neq 0}^{JN} |\tilde b_i|^2/\sigma_i^2 > NP$ or $\tilde b_i\neq 0$ for some $i$ such that $\sigma_i=0$, then $\tilde s_i= \tilde b_i/(\sigma_i+\lambda)$, for all $i$, where $\lambda >0$ is the unique solution of $f(\lambda)=NP$.
 \item Otherwise, $\tilde s_i= \tilde b_i/\sigma_i$ for all $i$ such that $\sigma_i\neq 0$, and $\tilde s_i= 0$, for all $i$ such that $\sigma_i= 0$.
\end{itemize}

Notice that upper and lower bounds to the optimal Lagrange multiplier $\lambda$ obtained from the solution to $f(\lambda)=NP$ are available (see also~\cite{Hager_2001}). Indeed, since
\begin{equation}
 NP= \sum_{i=1}^{JN} \frac{|\tilde b_i|^2}{(\sigma_i+\lambda)^2}  \leq \sum_{i=1}^{JN} \frac{|\tilde b_i|^2}{(\sigma_\text{min}+\lambda)^2} =\frac{\Vert \bm b \Vert^2}{(\sigma_\text{min}+\lambda)^2},
\end{equation}
where $\sigma_\text{min}=\min_{i=1,\ldots,JN} \sigma_i$, we have that
\begin{equation}
\lambda \leq \frac{\Vert \bm b \Vert}{\sqrt{NP}} - \sigma_\text{min}. \label{lambda_ub}
\end{equation}
Similarly, since
\begin{equation}
 NP= \sum_{i=1}^{JN} \frac{|\tilde b_i|^2}{(\sigma_i+\lambda)^2} \geq \frac{1}{(\sigma_\text{min}+\lambda)^2} \sum_{i=1, i : \sigma_i= \sigma_\text{min}}^{JN} |\tilde b_i|^2,
\end{equation}
we have that
\begin{equation}
 \lambda \geq  \max\left\{ 0,\sqrt{\frac{1}{NP} \sum_{i=1, i : \sigma_i= \sigma_\text{min}}^{JN} |\tilde b_i|^2} - \sigma_\text{min}\right\}. \label{lambda_lb}
\end{equation}
Therefore, the solution to $f(\lambda)=NP$ can easily be found using the bounds in~\eqref{lambda_ub} and~\eqref{lambda_lb} and exploiting the fact that $f(\lambda)$ is continuous, strictly decreasing, and strictly convex for $\lambda>0$ (e.g., we can use the bisection algorithm).

\bibliographystyle{IEEEtran} 

\end{document}